\documentclass[journal,twoside, web]{ieeecolor}
\usepackage{jsen}
\usepackage{cite}
\usepackage{amsmath,amssymb,amsfonts}
\usepackage{algorithmic}
\usepackage{graphicx}
\usepackage{textcomp}
\usepackage{wrapfig}
\usepackage{csquotes}
\usepackage{bm}
\def\BibTeX{{\rm B\kern-.05em{\sc i\kern-.025em b}\kern-.08em
    T\kern-.1667em\lower.7ex\hbox{E}\kern-.125emX}}
\definecolor{abstractbg}{rgb}{0.89804,0.94510,0.83137}
\begin{document}

\title{Generative Adversarial Reconstruction with Adaptive Thresholding for Obstructed Targets in Computational Microwave Imaging}
\author{Jiaming~Zhang,~\IEEEmembership{Graduate Student Member,~IEEE,}
        María García-Fernández,~\IEEEmembership{Senior Member,~IEEE,}
        Guillermo Álvarez-Narciandi,~\IEEEmembership{Senior Member,~IEEE,}
        Jie Zhang,~\IEEEmembership{Senior Member,~IEEE,}
        Muhammad Ali Babar Abbasi,~\IEEEmembership{Senior Member,~IEEE,}
        and~Okan~Yurduseven,~\IEEEmembership{Senior Member,~IEEE}
\thanks{This work was supported by the Leverhulme Trust under Research Leadership Award RL-2019-019, and by the UKRI Horizon Europe Guarantee for Marie Skłodowska-Curie Action Postdoctoral Fellowships under Projects EP/X022951/1 and EP/X022943/1. (\textit{Corresponding author: Jiaming Zhang.})}
\thanks{Jiaming Zhang, Jie Zhang, Muhammad Ali Babar Abbasi and Okan Yurduseven are with the Centre for Wireless Innovation, Queen's University Belfast, Belfast, United Kingdom, BT3 9DT. (e-mail: jzhang57@qub.ac.uk).
María García-Fernández, Guillermo Álvarez-Narciandi are with Group of Signal Theory and Communications, Department of Electrical, Electronic, Communications and Systems Engineering, University of Oviedo, 33203
Gijón, Spain.}
}

\IEEEtitleabstractindextext{%
\fcolorbox{abstractbg}{abstractbg}{%
\begin{minipage}{\textwidth}%
\begin{wrapfigure}[12]{r}{2.5in}%
\includegraphics[width=2.5in]{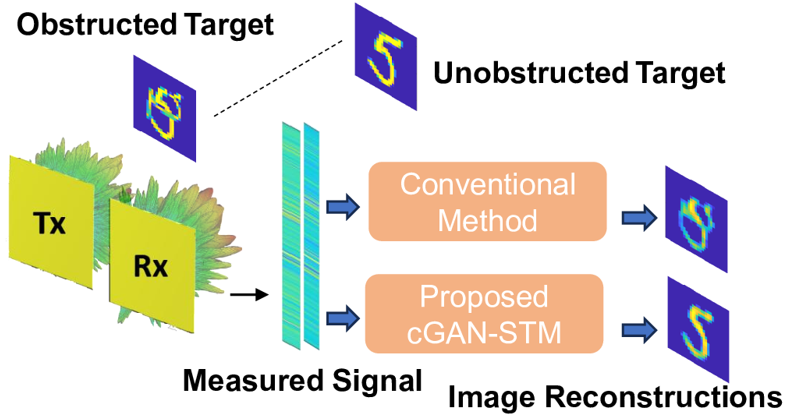}%
\end{wrapfigure}%
\begin{abstract}
In this work, an end-to-end generative adversarial framework for computational microwave imaging (CMI) is proposed to reconstruct the targets of interest directly from the measurements of targets obstructed by undesired objects. It integrates a conditional generative adversarial network (cGAN) with a learnable soft-threshold module (STM) to adaptively suppress non-target related information. The proposed framework is evaluated on a diverse dataset comprising MNIST digits obstructed by E-MNIST letters for training and testing, as well as on measurements acquired with an experimental CMI system. In addition, further studies involving objects with different geometries are conducted, demonstrating that the proposed approach can be adapted to other types of objects. Numerical experiments show that the proposed cGAN-STM achieves a normalized mean square error (NMSE) of 0.066 and a structural similarity index (SSIM) of 0.876 under ideal conditions. Comprehensive analyses, including benchmarking, analysis of the STM mechanism, and evaluation under different obstruction sizes, are also conducted. The performance of the model under different signal-to-noise ratio (SNR) scenarios is also evaluated, achieving reasonable reconstruction quality at 15 dB SNR with an NMSE of 0.168 and an SSIM of 0.700. Even at low SNR levels, recognizable target outlines are preserved. These results highlight the effectiveness and adaptability of the proposed method.
\end{abstract}

\end{minipage}}}

\maketitle

\section{Introduction}
\label{sec: Introduction}
\subsection{Background}
Microwave imaging has gained significant attention across a wide range of applications, including localization, debris detection, and earth science, to name a few \cite{yurduseven2019frequency,he2023framework, 10103561,yurdusevne2020frequency,9108406}. Conventional microwave imaging techniques mainly rely on synthetic aperture radar (SAR)-based modalities \cite{10521717,10384674,fromenteze2019transverse}. In SAR-based approaches, the transmitters (Tx) and receivers (Rx) are either mechanically or electronically scanned at the Nyquist sampling interval, leading to a point-by-point interrogation (raster scan) of the imaging scene \cite{6504845,10537210}. Although mechanically scanned SAR systems can provide high-fidelity reconstructions, they suffer from time-consuming data acquisition due to sequential antenna motion \cite{bolomey2008overview, 10115470,chen2021real}. On the other hand, electronically scanned array systems \cite{mailloux2007electronically,hansen2009phased} overcome this limitation by enabling fast beam steering, but at the cost of intensive hardware complexity and increased power consumption.

To tackle these issues, computational microwave imaging (CMI)-based systems \cite{molaei2025advanced} are explored as an alternative. Such systems employ a set of spatially incoherent radiation patterns (or measurement modes) to probe the scene under investigation, achieving a physical layer compression \cite{10415384,8972939, doi:10.1126/science.1230054}. CMI-based systems can be categorized by how the spatially-incoherent radiation patterns are generated: dynamic reconfiguration of the aperture \cite{doi:10.1063/1.4935941,10564005,yurduseven2017millimeter} or frequency diversity \cite{10472621,10539934,hoang2021spatial}. In particular, dynamically reconfigured CMI-based systems use a tuning mechanism to actively modify the aperture of the antenna. As a result, a diverse set of radiation patterns can be generated, eliminating the need for large bandwidths \cite{9779101}. On the other hand, frequency-diverse CMI-based systems require a frequency bandwidth of operation to generate frequency-distinct radiation patterns by utilizing passive structures. While CMI significantly reduces the hardware complexity by decreasing the number of data acquisition channels, the computational burden of the image reconstruction step can hinder its deployment in real-world applications \cite{7557020}. 

\subsection{Related Work}
Motivated by the need to reduce the computational burden while maintaining reconstruction quality, recent research has increasingly adopted deep learning approaches for microwave imaging \cite{9852109,10974998,Ria_Benny}. For instance, a novel U-Net was proposed in \cite{9107447} to reconstruct images from phaseless microwave data. {\color{black} \cite{10352954} also proposed a U-Net–integrated reconstruction scheme that improves the accuracy of dielectric constant estimation for uniaxial objects in microwave imaging.} Similarly, in \cite{9970516}, an integrated deep learning system was designed to predict multi-frequency electromagnetic (EM) scattered fields from single-frequency EM scattered fields and to generate the corresponding image reconstructions. Additionally, a two-stage training method was developed in \cite{9034483}, significantly reducing the difficulty of training deep neural networks for image reconstruction. {\color{black}A physics-guided reconstruction approach was proposed in \cite{10214437} to recover fine-scale permittivity distributions from limited and noisy scattered-field measurements.}

Standard feed-forward neural networks proposed in the above literature aim to minimize the loss function that evaluates the quality of results. However, this minimization often leads to solutions that average all plausible element values, resulting in poor perceptual quality \cite{8099502, NIPS2016_371bce7d}. To address this issue, generative adversarial networks (GANs) are employed to produce outputs with high perceptual quality \cite{goodfellow2020generative, 8100115}. In particular, conditional GANs (cGANs) are widely applied to solve image reconstruction problems. For example, in \cite{9676480}, an attention-assisted cGAN was designed for the image reconstruction problem with mixed boundary conditions. This approach eliminated the need for prior knowledge of the boundary conditions when reconstructing the image of the scene. Additionally, \cite{10473754} applied a deep convolutional cGAN, termed DCCGAN, to transform the single-transmitter-single-frequency measured signals into the reconstructions of the scene. In \cite{9996181}, a novel cGAN was developed with self-attention modules and trained with a novel weighted loss function. This network was capable of solving highly nonlinear inverse scattering problems. {\color{black}\cite{9900447} developed a deep unfolding GAN within the contrast source inversion framework to improve image reconstruction quality under challenging imaging conditions.}

To address different CMI tasks simultaneously, our previous work proposed a deep learning model named ClassiGAN to concurrently solve the image reconstruction and target classification tasks by only using the back-scattered measured signals as input \cite{10892224}. Also, in \cite{10380631}, a diffusion model was proposed to achieve image reconstructions based on the scattering signal. However, compared to cGAN-based methods, diffusion models require a large number of iterative denoising steps \cite{ho2020denoising}, which substantially reduces their computational efficiency.

\subsection{Motivation and Contribution}
A common feature of the previous studies highlighted earlier is that they consider EM imaging scenarios where the imaged objects are present without any obstructions. This can pose a significant limitation for practical applications as, in reality, an imaged target can be obstructed by secondary objects that can significantly impact the back-scattered information captured from the scene of interest for imaging. In addition to significantly impacting the accuracy of the reconstructed images, such obstructions can also necessitate manual intervention to physically separate the target information from the obstructing objects, which can be time consuming, inefficient, and even impractical due to high false-alarm rates \cite{horsley2019microwave, 6305002}. 

To tackle the aforementioned issues, in this work, a novel cGAN integrated with deep learning-based soft-threshold modules (STMs) is designed. The proposed technique, termed cGAN-STM, significantly reduces the error caused by the obstruction and learns the relevant features of targets from the CMI-measured signals, achieving high-quality reconstructions of clean targets. To the best of the authors' knowledge, this work {\color{black}presents a proof-of-concept implementation of a deep generative adversarial learning framework for image reconstruction with obstructed imaging targets in CMI.} The main contributions are summarized below:

\begin{itemize} 

\item Design of the cGAN-STM, which is capable of transforming the measured signals directly into the reconstructions of the obstructed imaging targets of interest. Compared to traditional CMI methods, this approach reduces the computational cost and eliminates the need for manually removing obstructing objects prior to accurate reconstructions.

\item Development of a deep learning-based STM to filter out the information associated with the obstructing objects and retrieve that corresponding to the targets of interest. Instead of manually selecting the threshold values, which can slow down the process, the optimal values are learned by this module. 

\item Benchmark against existing state-of-the-art deep learning methods showing that cGAN-STM achieves superior quantitative and qualitative performance in reconstructing obstructed targets.

\item {\color{black}The proposed framework was extensively tested through synthetic data and using measurements obtained with an experimental CMI system. The use of controlled synthetic conditions enabled the assessment of the impact of varying obstruction sizes, STM ablation, different noise conditions, adaptability to unseen targets and STM mechanism analysis. The results demonstrate that the proposed framework effectively reconstructs obstructed targets and exhibits robustness and adaptability across different setups, supporting its potential scalability toward more complex CMI scenarios.}

\end{itemize}

The remainder of this article is organized as follows. Section \ref{sec: Computational Microwave Imaging} introduces the CMI paradigm and the image reconstruction algorithms. Section \ref{sec: Proposed method} introduces and explains the functionality of cGAN-STM. The results are shown and discussed in Section \ref{sec: Results and Discussion}. Finally, the main conclusions are drawn in Section \ref{sec: Conclusion}. 

\section{Computational Microwave Imaging}
\label{sec: Computational Microwave Imaging}

\begin{figure}[t]
\centering
    \includegraphics[width=\columnwidth]{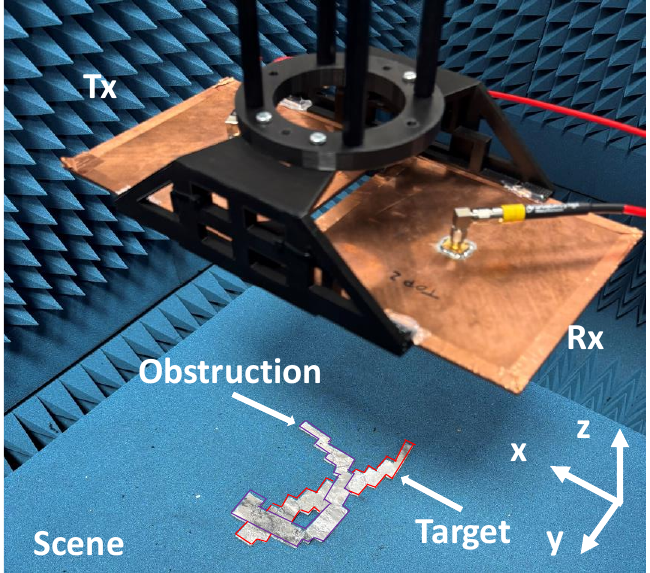} 
    \caption{Bi-static CMI setup considered for the data generation and for experimental verification. The target example includes an obstruction outlined in purple and a target outlined in red. Both materials are aluminate.} 
    \label{fig:antenna}
\end{figure}

A CMI-based system probes the imaged scene by using {\color{black}quasi-random spatially incoherent radiation patterns. The term \enquote{quasi-random} highlights that the patterns are not purely stochastic but arise from the specific structure or tuning of the metasurface} {\color{black}\cite{8972939}},{\color{black}\cite{10178101, 10944265}. Meanwhile, \enquote{spatially incoherent} indicates that different modes are nearly orthogonal, facilitating that, ideally, each illumination probes the scene independently and thus enables efficient information encoding {\color{black}\cite{8972939}}.} The scattered field \textit{$E_{\mathrm{scat}}$}, which contains the reflectivity distribution of the scene, ${{\rho} (\bm{r})}$ \cite{Pulido-Mancera:16}, according to the first Born approximation \cite{doi:10.1126/science.1230054},{\color{black}\cite{7927481, 9650546}}, is given as:
\begin{equation}
    E_{\mathrm{scat}} = \int_{V} {\rho}(\bm{r}) E_{\mathrm{inc}} dV ,
\end{equation}
where $E_{\mathrm{inc}}$ is the incident field from the aperture and $\bm{r}$ denotes the coordinates vector of the scene. In this work, the bold font is used to denote the vector-matrix notation. For numerical implementation, the scene is discretized into a finite number of pixels, such that the reflectivity distribution, $\boldsymbol{\rho}$, can be represented in vector form. The transfer function of the CMI system, $\mathbf{H}$, which is also called the sensing matrix, can be derived by the dot product of fields radiated by the transmit and receive apertures forming the CMI system, $\mathbf{E}_{Tx}$ and $\mathbf{E}_{Rx}$ respectively, both of which are propagated to the imaging scene \cite{lipworth2015comprehensive}:
\begin{equation}
   \mathbf{H}=\mathbf{E}_{Tx} \cdot \mathbf{E}_{Rx}.
   \label{eqH}
\end{equation}

The back-scattered measured signal $\mathbf{g}$ can be obtained by:
\begin{equation}
    \mathbf{g}_{M\times 1}= \mathbf{H}_{M\times N}\boldsymbol{\rho}_{N\times 1}+\mathbf{n}_{M\times1},
    \label{eq1}
\end{equation}
where $\mathbf{g}$ is a vector of length $M$, while $\boldsymbol{\rho}$ is a vector of length $N$, corresponding to the reflectivity of the scene. Here, $M$ denotes the number of measurement modes, while $N$ denotes the number of pixels of the imaging scene.  It should be noted that the sensing matrix is not necessarily square, i.e., $M \neq N$. Additionally, $\mathbf{n}$ is a noise term, which is a vector of length $M$. 

In this study, the conventional CMI reconstructions are obtained using a commonly employed approach, the least-squares technique given in (\ref{eq2_ls}), which retrieves an estimate of the reflectivity of the scene ($\boldsymbol{\rho}_{\textrm{rec}}^{\textrm{LS}}$), and they are compared against the proposed cGAN-STM predicted reconstructions (estimated directly from g without reconstructing $\boldsymbol{\rho}_{\textrm{rec}}$).
\begin{equation}
    \boldsymbol{\rho}_{\textrm{rec}}^{\textrm{LS}}= \mathop{\arg \min}_{\boldsymbol{\rho}}\lvert \lvert \mathbf{g} - \mathbf{H}\boldsymbol{\rho} \rvert \rvert_{2}^{2}
    \label{eq2_ls}
\end{equation}

Here, $\lvert \lvert \cdot \rvert \rvert_{2}$ indicates the L2-norm. The CMI-based system synthesized in this study relies on the two-dimensional (2D) frequency-diverse metasurface antenna presented in \cite{10486957}. It comprises two frequency-diverse antennas (or panels), one acting as a Tx and the other as a Rx arranged side by side, operating in a bi-static operating configuration, as shown in Fig. \ref{fig:antenna}. In this work, frequency-diverse antennas operating in the X-band (8–12 GHz) are employed, with further design details provided in \cite{10486957}. To obtain high-resolution reconstructions, a hybrid CMI-SAR approach \cite{8055576} is employed to collect sufficient information on the scene. Specifically, the measurement grid consists of 16 positions, spaced 8 cm apart along both the \textit{x}-axis and \textit{y}-axis. The total number of measurement modes, $M$, is 1024, and the number of pixels in the scene, $N$, is 784.

\section{Proposed Approach}
\label{sec: Proposed method}
\subsection{Motivation}

In an imaging scenario with obstructed targets, the back-scattered measurements contain information from both the targets of interest and the obstructing objects. Since the task in this work is to directly reconstruct the targets of interest, the information related to the obstructing objects can be regarded as task-specific noise. To mitigate the degradation of image quality caused by obstructions, an automatically learned STM is proposed. {\color{black}The STM adaptively suppresses non-target features while preserving relevant target information}, guiding the network to focus on meaningful signals. As will be shown later, the proposed STM plays a crucial role in ensuring robust performance against obstruction-related artifacts.

To predict the missing target information in obscured regions, generative adversarial learning is employed, enabling the network to infer obstructed parts from the measurements of occluded targets and to generate high-fidelity reconstructions. The STM and adversarial learning modules act in a complementary manner, ensuring accurate and robust reconstruction in obstructed scenarios.

\subsection{Soft-thresholding Module (STM) Architecture}
\label{subsec:STMArchi}
To enhance the network's ability to separate the target information from the interference caused by obstructions, a soft-threshold mechanism is employed. The soft-threshold function is defined as:
\begin{equation}
V_o =
\begin{cases}
V_i - \tau, & \text{if } V_i > \tau, \\
0, & \text{if } -\tau \leq V_i \leq \tau, \\
V_i + \tau, & \text{if } V_i < -\tau,
\end{cases}
\label{eq:st}
\end{equation}
where $V_i$ and $V_o$ respectively correspond to the input and output data, and $\tau$ denotes the threshold value. 

Although (\ref{eq:st}) shares the same mathematical form as the shrinkage operator commonly used in classical compressed sensing \cite{woodworth2016compressed}, the soft-threshold operation applied in the cGAN feature domain serves as an adaptive filter that suppresses obstruction-induced activations instead of promoting sparsity in the reconstructed image or measurement domain.

Equation (\ref{eq:st}) shows that the soft-threshold mechanism is able to highlight salient features by retaining strong activations as positive or negative values, while suppressing weak, potentially redundant activations toward zero. However, determining the appropriate threshold value $\tau$ typically relies on expert knowledge or an optimization process, which can be time-consuming \cite{8850096}. Therefore, inspired by the residual shrinkage building unit in \cite{8850096}, a sub-network called STM is designed to automatically learn the optimal threshold values, eliminating the need for manual threshold selection. 

Fig. \ref{fig:network}(a) illustrates the architecture of the STM. As shown in Fig. \ref{fig:network}(a), a global average pooling layer \cite{Lin2014} is first applied to process the absolute values of the input feature maps. The global average pooled results are then passed through two fully connected layers. The first fully connected layer is followed by a batch normalization (BatchNorm) layer and a rectified linear unit (ReLU) activation function. The second fully connected layer is followed by a sigmoid activation function \cite{10658992}, which scales the output to the range (0,1). Finally, the outputs of the fully connected layers are element-wise multiplied with the globally averaged features to determine the threshold values $\tau$. The STM output is computed by applying these calculated threshold values in conjunction with the soft-threshold function defined in (\ref{eq:st}). The learnable thresholds $\tau$ generated by the sub-network are differentiable and can be updated through back-propagation, enabling the STM to adaptively adjust its filtering strength to different inputs during training.

\subsection{Network Architecture}
\label{sec:Network}
\begin{figure*}[h]
    \centering
    \includegraphics[width=\textwidth]{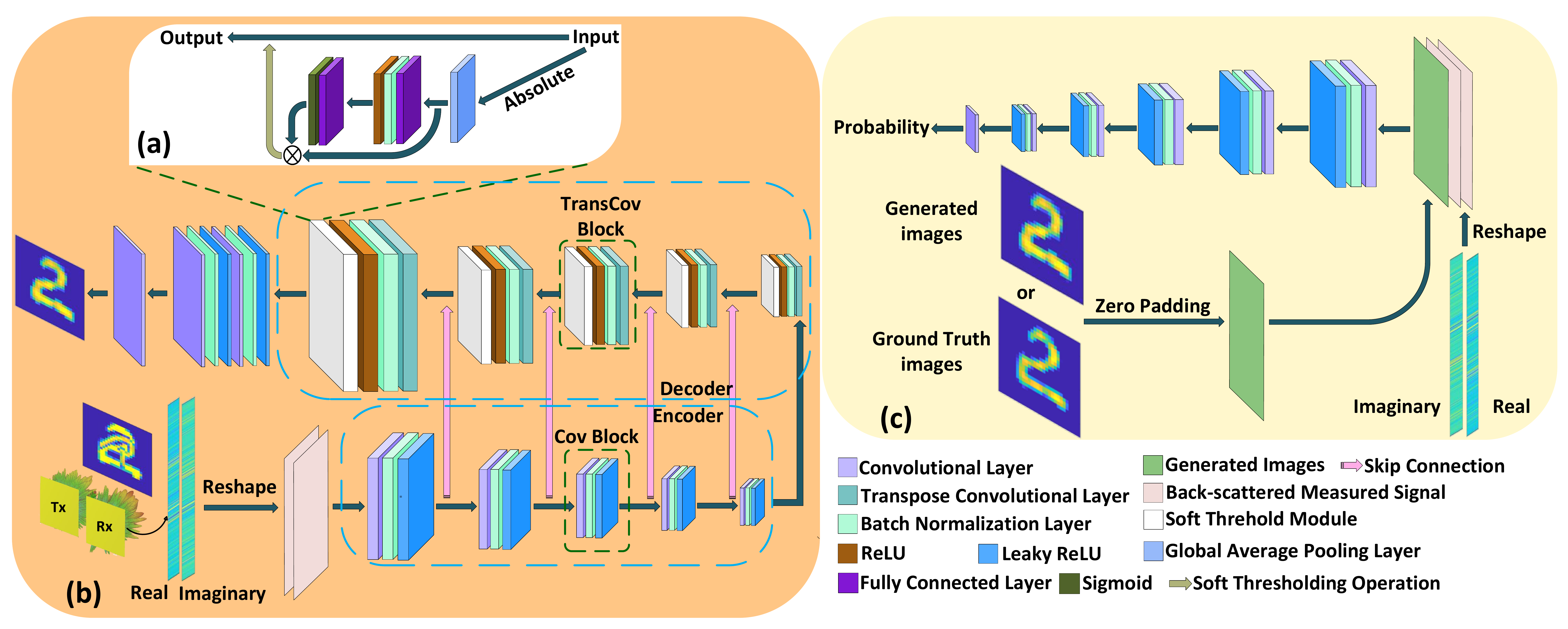}
    \caption{Overall network architecture highlighting the obstruction suppression and target reconstruction modules: (a) STM for adaptive feature filtering, (b) generator with encoder-decoder for feature extraction and reconstruction, (c) discriminator for adversarial quality enforcement.}
    \label{fig:network}
\end{figure*}
cGAN-STM comprises two main components: a generator and a discriminator. The generator is used to transform the back-scattered measurements into image reconstructions, while the discriminator is applied to distinguish the authenticity of the image reconstructions generated by the generator. They are trained in an adversarial manner. 

The detailed architecture of the generator is shown in Fig. \ref{fig:network}(b). It mainly uses a U-Net architecture consisting of an encoder and a decoder. The encoder compresses the complex-valued back-scattered measurements into latent feature representations that emphasize target-related information. The decoder subsequently reconstructs the targets by refining these representations and recovering spatial details suppressed by obstructions. The main element of the encoder is the convolutional block (Conv Block), which includes a convolutional layer, a BatchNorm layer, and a leaky ReLU activation function \cite{9163446}. Similarly, the decoder's primary component is the transpose convolutional block (TransConv Block) composed of a transpose convolutional layer \cite{Long_2015_CVPR}, a BatchNorm layer, a ReLU activation function and an STM. The STM employed in the decoder aids in retaining essential features while reducing redundant data, thereby minimizing the influence of obstruction-induced interference during the processing of compressed information. To mitigate information loss throughout the encoding and decoding stages, skip connections are end-to-end connected between the corresponding Conv Blocks and TransConv Blocks \cite{he2016identity, dai2023integrating}. As discussed before, the number of measurement modes in the back-scattered data is $M = 1024$, which exceeds the number of pixels in the target image reconstructions, $N=784$. After being processed by the symmetric U-Net, the decoder output matches the size of the encoder input rather than the desired image resolution. Therefore, two additional convolutional blocks are applied to reduce the decoder output to the target image size of 28×28 pixels, ensuring consistency for effective training.

The discriminator is shown in Fig. \ref{fig:network}(c). It shares the same architecture as the encoder of the generator, with the addition of a convolutional layer with a single kernel to provide the probability of the image reconstruction's authenticity \cite{9862978}. In a similar fashion as in the input of the encoder of U-Net in the generator, the complex-valued back-scattered measurements are separated into two channels and reshaped into a square format. Since the size of the images is smaller than that of the reshaped backscattered measurements, the zero-padding technique is applied to enlarge the images. This allows the images to be concatenated with the reshaped backscattered measurements. The concatenated input data is then fed into the discriminator to judge whether the inputted images are predicted reconstructed images or the ground truth reconstructed images, thereby enforcing the generator to produce reconstructions that are visually realistic and free from obstruction.

\subsection{Loss Functions}
\label{sec:loss}
Let $\mathcal{L}_{D}$ denote the adversarial loss for the discriminator. This loss is computed using the binary cross-entropy (BCE) function. Thus, the general form of $\mathcal{L}_{D}$ is defined as:

\begin{equation}
\mathcal{L}_{D} = -\mathbb{E}_{x_i} \left[ y_i \log_{e}(x_i) \right] -\mathbb{E}_{x_i}\left[(1-y_i)\log_{e}(1-x_i)\right],
\end{equation}
where $\mathbb{E}$ denotes the expectation, $x_i$ represents the discriminator’s predicted probability that the input data is actual, and $y_i$ is the corresponding ground-truth label. A value of $x_i$ closer to 1 indicates higher realism, while a value closer to 0 indicates the data is likely generated.

Let $\mathcal{L}_{D}(\mathrm{Generated})$ and $\mathcal{L}_{D}(\mathrm{Actual})$ denote the discriminator loss terms corresponding to generated and ground-truth image reconstructions, respectively. To guide the discriminator in distinguishing actual from generated image reconstructions, $y_i$ is set to 0 for generated image reconstructions and 1 for actual (ground-truth) image reconstructions. $\mathcal{L}_{D}(\mathrm{Actual})$ is thus given by:
{\color{black}
\begin{equation} 
    \begin{split}
    \mathcal{L}_{D}(\mathrm{Actual}) 
    ={}& -\mathbb{E}_{\mathbf{g},R}[\log (D(\mathbf{g},R))],
    \end{split}
\label{eq_lossD_Dis}
\end{equation}  }              
where $R$ is the ground-truth reconstructed image, and $D(\cdot)$ outputs the discriminator’s estimated probability that the image is actual. Similarly, $\mathcal{L}_{D}(\mathrm{Generated})$ is given by:
{\color{black}
\begin{equation} 
    \begin{split}
    \mathcal{L}_{D}(\mathrm{Generated})  
    ={}&-\mathbb{E}_{\mathbf{g},G(\mathbf{\mathbf{g}})}[(\log (1-D(\mathbf{g},G(\mathbf{g}))))],\\  
    \end{split}
\label{eq_lossD_Gen}
\end{equation} }
where $G(\mathbf{g})$ indicates the predicted image reconstructions. Combining (\ref{eq_lossD_Dis}) and (\ref{eq_lossD_Gen}), the discriminator loss function is expressed as follows:
\begin{equation} 
    \begin{split}
    \mathcal{L}_{D} = {}& \mathcal{L}_{D}(\mathrm{Generated})+\mathcal{L}_{D}(\mathrm{Actual})\\
    ={}&-\mathbb{E}_{\mathbf{g},G(\mathbf{\mathbf{g}})}[(\log (1-D(\mathbf{g},G(\mathbf{g}))))]\\
     {}& -\mathbb{E}_{\mathbf{g},R}[\log (D(\mathbf{g},R))],\\
    \end{split}
\label{eq_lossD}
\end{equation}

For the generator loss function, since the generator aims to \enquote{fool} the discriminator so that the generated image reconstruction is identified to be the ground-truth, $y_i$ should be one. Denoting $\mathcal{L}'_{D}(\mathrm{Generated})$ as the adversarial loss term for the training of the generator, $\mathcal{L}'_{D}(\mathrm{Generated})$ can be computed as:

\begin{equation} 
    \begin{split}
    \mathcal{L}'_{D}(\mathrm{Generated})  
    ={}&-\mathbb{E}_{\mathbf{g},G(\mathbf{\mathbf{g}})}[D(\mathbf{g},G(\mathbf{g}))))]\\
    \end{split}
\label{eq_lossG_Dis}
\end{equation}

Therefore, the total loss function for the generator is given by:
\begin{equation} 
    \begin{split}
    \mathcal{L}_{G} = {}& \lambda \mathcal{L}_{L1} + \mathcal{L}'_{D}(\mathrm{\mathrm{Generated}}) \\
    ={}&\lambda \mathbb{E}_{R,G(\mathbf{g})}[\lvert \lvert R - G(\mathbf{g}) \rvert \rvert_{1}]\\ 
    {}&- \mathbb{E}_{\mathbf{g},G(\mathbf{g})}[\log (D(\mathbf{g},G(\mathbf{g})))],\\
    \end{split}
\label{eq_lossG}
\end{equation}
where $\lvert \lvert \cdot \rvert \rvert_{1}$ denotes the L1-Norm. Employing the L1-Norm can enhance the pixel-wise similarity between the generated output and the ground truth. {\color{black}In addition, consistent with the Pix2Pix framework, the coefficient $\lambda$ in (\ref{eq_lossG}) is set to 100. Several studies have empirically shown this choice to provide a good balance between the adversarial loss and the L1-Norm loss, and thus yields stable and high-quality results }\cite{10130328,10505767,8100115}.


\subsection{Back-scattered Data and Image Reconstruction Generation}
\label{sec: Data Generation}
Firstly, the training and testing datasets consist of synthetic back-scattered measurement signals and their related image reconstructions. {\color{black}The training dataset is used to learn the general mapping between back-scattered measurements and unobstructed reconstructions, while the testing dataset, consisting of distinct samples from the same distribution, is used to evaluate the model’s performance.}

To generate the datasets, the fields radiated by the transmit and receive antennas are experimentally characterized using a planar antenna measurement range, calculating the sensing matrix as described in \cite{10486957}. Then, the back-scattered measurements of obstructed targets are subsequently calculated by employing the reflectivity of these targets and using (\ref{eq1}). Finally, the image reconstructions of these targets are obtained by employing (\ref{eq2_ls}). To obtain obstruction-free image reconstructions, the same steps are implemented on the unobstructed targets of interest.

Each sample set in the datasets consists of the back-scattered measurement of an obstructed target and the image reconstruction of the related target of interest without any obstructions, which corresponds to the input and the output of the network, respectively. The targets of interest considered in this paper are handwritten digits from the open-source dataset MNIST \cite{726791}, and the obstructing objects are hand-written letters from the open-sourced dataset E-MNIST \cite{7966217}. Each MNIST or E-MNIST image is first converted into a two-dimensional reflectivity distribution by normalizing the pixel values to the range [0,1], such that higher intensity pixels correspond to stronger reflectivity. This reflectivity map is then employed as the target model in the forward model (\ref{eq1}), together with the experimentally characterized sensing matrix, to generate the synthetic back-scattered measurements.

\begin{figure}[htb]
    \centering
    \includegraphics[width=\columnwidth]{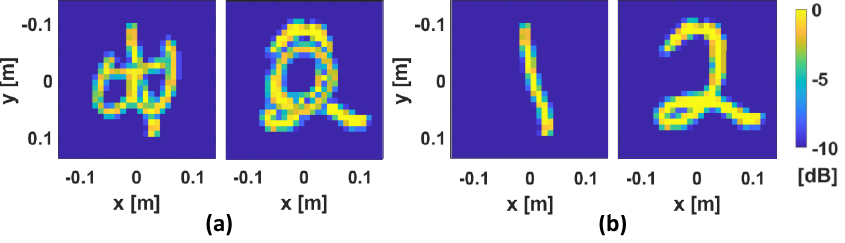}
    \caption{Examples of: (a) obstructed targets and (b) the corresponding targets of interest (unobstructed). {\color{black}The colour scale represents normalized reflectivity in dB. The maximum (1) is displayed at $0$ dB, and values below $-10$ dB are clipped.}}
    \label{fig:targets}
\end{figure}

Employing MNIST as imaging targets of interest and E-MNIST as obstructions is motivated by their great variability due to the natural differences in handwriting styles. In addition, the obstructing letters are randomly selected, randomly positioned, and scaled relative to the target digits, creating diverse occlusion patterns across the dataset. This combination of inherent handwriting variability and randomized occlusions enhances the diversity of the back-scattered measurements and image reconstructions, contributing to the robustness and generalization capability of cGAN-STM to different partial occlusion scenarios. 

When generating each obstructed target, a randomly selected obstructing object (from the E-MNIST dataset) is placed in front of a randomly chosen target of interest (from the MNIST dataset), resulting in an overlap that covers the target of interest. {\color{black}Both the obstructing object and the target are assumed to lie on the same imaging plane, corresponding to two thin layers in close contact with negligible separation (as illustrated by the target used for experimental validation in Fig. \ref{fig:antenna}). This modelling simplification enables the obstacle to directly occlude the target within the 2D framework.} In addition, to provide enough information for the targets of interest (digits), the obstructing objects (letters) are scaled to 60\% of the targets of interest size. The impact of varying the size of the obstructing objects will be investigated later in Section \ref{sec:size}. Figs. \ref{fig:targets}(a) and  \ref{fig:targets}(b) respectively show two examples of the obstructed targets and their related targets of interest employed for the generation of the training and testing dataset. For visualization in the logarithmic (dB) scale, the maximum value (corresponding to 1) is displayed at 0 dB, while the lower bound of the dynamic range is clipped to -10 dB for display purposes.

Note that MNIST and E-MNIST are both publicly divided into the training and testing subsets. To guarantee the distinctiveness between the training and testing datasets employed for this work, the obstructed targets associated with training sample sets comprise the digits and letters from the training sub-sets of MNIST and E-MNIST, respectively. Similarly, the testing sample sets are generated using the obstructed targets consisting of the digits from the testing sub-sets of MNIST and letters from the testing sub-sets of E-MNIST. 180000 sample sets and 10000 sample sets are defined as the training dataset and the testing dataset, respectively. 

\subsection{Implementation details}
Given that these back-scattered measurements are complex-valued, the magnitude and phase components are separated into two channels before being fed into the generator. Therefore, the size of the input data is 1024 $\times$ 1 $\times$ 2. The input information is then reshaped into a square matrix format to align with the U-Net. Similarly, the size of the reconstructed images is 28 $\times$ 28 $\times$ 1, corresponding to the size of the scene.

For the generator, both the convolutional and transpose convolutional layers in the U-Net architecture use 64 kernels, each with a kernel size of 4 $\times$ 4 and a stride of 2. Moreover, the convolutional layers employed for reducing the data size use 16 kernels, each sized 3 $\times$ 3 and with a stride of 1. Finally, since the output of cGAN-STM is a single channel output, a convolutional layer with a kernel size of 1 $\times$ 1 is utilized as the last layer of the generator. Similarly, to compute the probability of the authenticity of the image reconstructions, a convolutional layer with a 4 $\times$ 4 kernel is also applied at the end of the discriminator. All kernels in cGAN-STM are initialized using the Xavier initialization method \cite{pmlr-v9-glorot10a}.

The Adam optimizer with a learning rate of $1 \times 10^{-4}$ is applied to optimize the proposed generative model. To speed up the training process, the input data is standardized following a normal distribution with a mean of zero and a variance of one \cite{10415384}.
The output data is normalized within the range of -1 to 1. An Intel\textsuperscript{\textregistered} Core\textsuperscript{TM} i7-1265U CPU and a CUDA platform with an NVIDIA Quadro RTX A40 GPU with 24 GB dedicated memory size are used for both the training and testing processes. The numerical data is generated by MATLAB 2022a, and the experiments are conducted using Tensorflow 2.6.0. Because the loss functions tend to be stable at the $100^{\text{th}}$ epoch, the number of training epochs is chosen to be 100. {\color{black}Training the proposed cGAN-STM required 7 hours and 48 minutes. }Unless otherwise specified, all experiments use the same hyper-parameters, trained once on the noise-free training dataset (Section~\ref{sec: Data Generation}), and no retraining is performed for subsequent evaluations.

\begin{figure}[h!]
    \centering
    \includegraphics[width=\columnwidth]{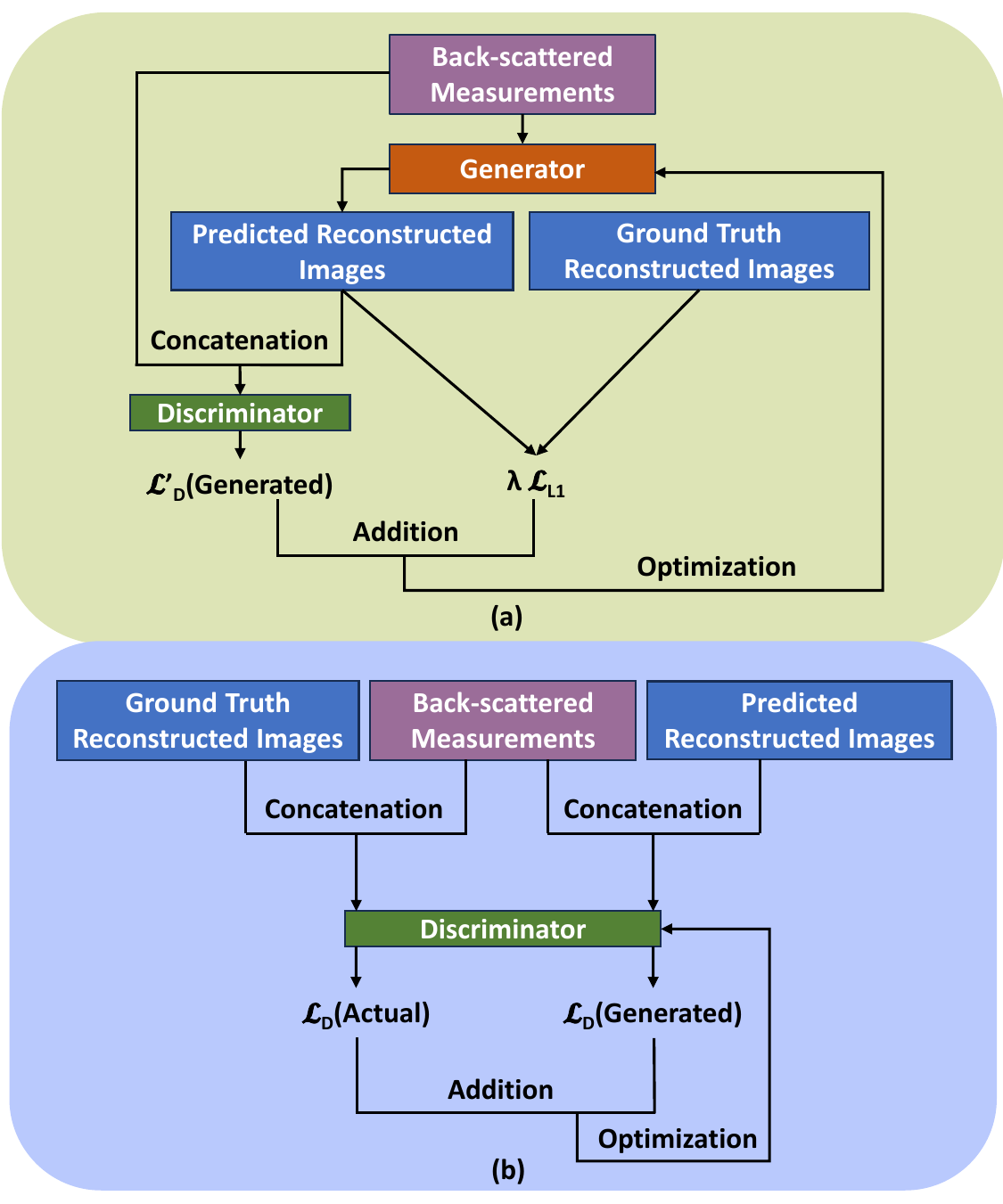}
    \caption{The training process of (a) the generator and (b) the discriminator.}
    \label{fig:trainingoverview}
\end{figure}

The training process of the generator and discriminator is shown in Fig. \ref{fig:trainingoverview}. As depicted in Fig. \ref{fig:trainingoverview}, the training process begins by feeding the back-scattered measurements of obstructed targets into the generator, which produces the corresponding reconstructed images. The generator’s parameters are updated based on the loss functions defined in (\ref{eq_lossG}). Following each update, the generator’s parameters are frozen, and the training proceeds to update the discriminator. During this phase, the discriminator is presented with pairs consisting of back-scattered measurements and either generated or ground-truth reconstructions. Its parameters are then optimized using the loss function specified in (\ref{eq_lossD}). This alternating optimization strategy continues iteratively until the network converges and delivers satisfactory reconstruction performance.

\section{Results and Discussion}
\label{sec: Results and Discussion}

\begin{figure*}[htb]
\centering
    \includegraphics[width=\textwidth]{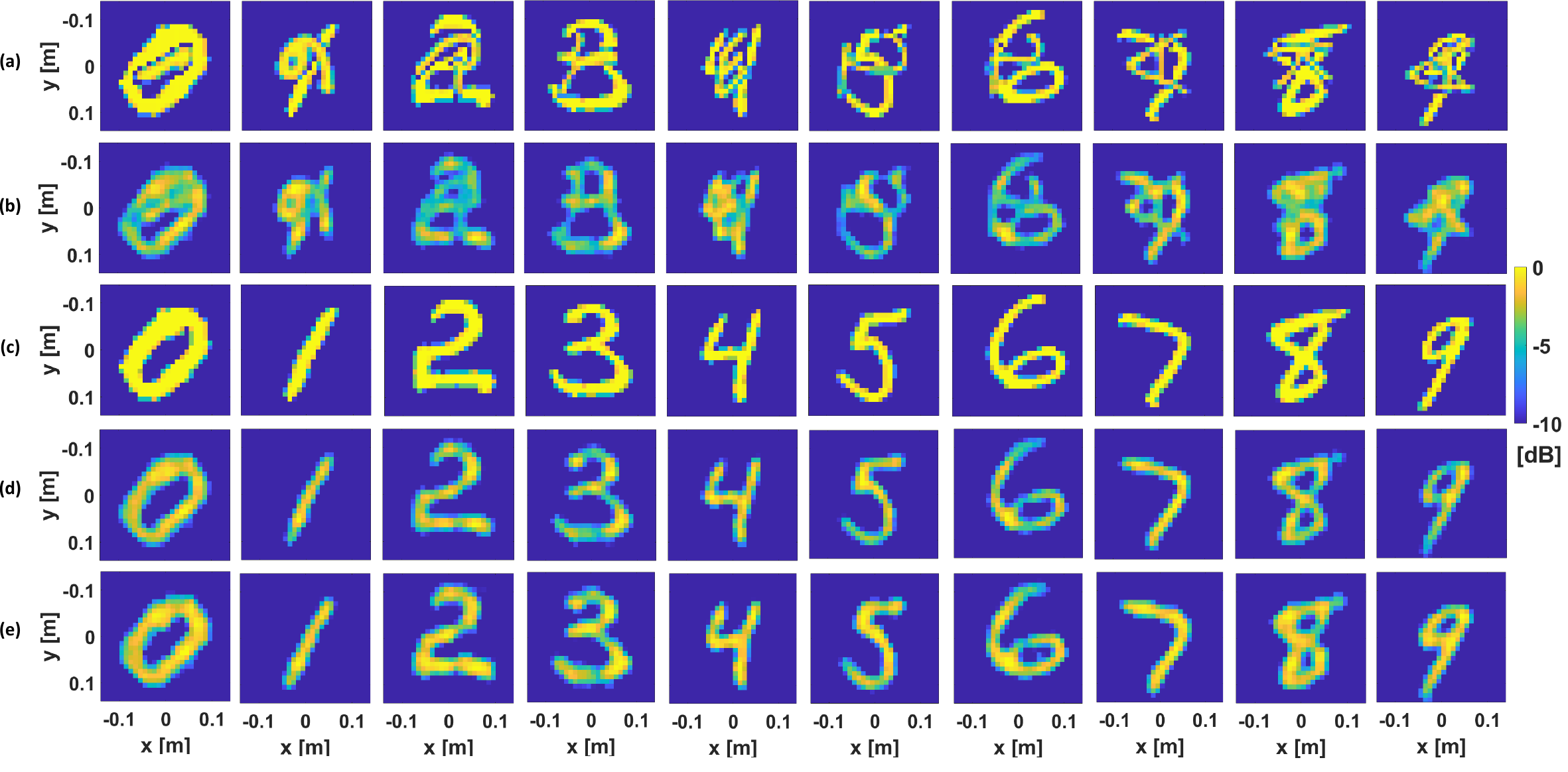} 
    \caption{Comparison of image reconstructions of the obstructed imaging targets based on numerically synthesized back-scattered measurements using the experimentally measured sensing matrix: (a) obstructed imaging targets; (b) the image reconstructions of (a) retrieved using the conventional method outlined in (\ref{eq2_ls}); (c) the targets of interest of (a); (d) the image reconstructions of (c) obtained by using the conventional method; and (e) the image reconstructions using the back-scattered measurements of obstructed targets shown in (a), obtained by the proposed cGAN-STM. {\color{black}The colour scale represents normalized reflectivity in dB. The maximum (1) is displayed at $0$ dB, and values below $-10$ dB are clipped.}} 
    \label{fig:numericResult1}
\end{figure*}

This section first introduces the evaluation metrics (Section~\ref{sec:Evaluation Metrics}), followed by the numerical and experimental results (Sections~\ref{sec:Numerical Evaluation} and \ref{sec:Experimental Evaluation})). The performance of the proposed cGAN-STM is further examined through an ablation study (Section~\ref{sec:ab}) and benchmarking comparisons with existing methods (Section~\ref{sec:Method Benchmarking}). The impacts of different obstruction sizes are analyzed in Sections~\ref{sec:Obstruction size1} and \ref{sec:size}. {\color{black}Section~\ref{Sec:Analysis of STM Mechanism} investigates the STM mechanism, while Section \ref{sec:Noise Study} presents a noise study. Section~\ref{sec:rev} analyzes occlusion suppression when both targets and obstructions are familiar to the model. Finally, Section~\ref{Sec:Adaptability Analysis} evaluates the adaptability of the proposed cGAN-STM, and Section~\ref{Sec:Limitation and Future Work} discusses its limitations and potential future work.}

\subsection{Evaluation Metrics}
\label{sec:Evaluation Metrics}
To evaluate the image reconstructions quantitatively, the normalized mean-squared-error (NMSE) \cite{9690176} and the structural similarity (SSIM) \cite{9905492} are used as the comparison metrics. In this work, the reconstructions obtained by the conventional imaging method in (\ref{eq2_ls}) are considered as the reference (or ground truth) for NMSE and SSIM calculations. In this way, the performance gap between the proposed cGAN-STM and the conventional method under identical measurement conditions can be assessed.

Specifically, the NMSE is used to calculate the pixel-wise accuracy between the ground truth reconstructed images generated by the conventional method, $\mathbf{R}$, and the predicted reconstructed images generated by the cGAN-STM and other state-of-the-art models for comparison, $\mathbf{F}$. The NMSE between them is given by:
\begin{equation}
    \mathrm{NMSE} = \frac{\frac{1}{N}\sum_{p=1}^{N}(R_{p} - F_{p})^{2}
    }{\frac{1}{N}\sum_{p=1}^{N}R_{p}^{2}},
\end{equation}
where the index $p$ denotes the pixel index. Additionally, the SSIM emphasizes the information of the generated images, including structural, luminance and contrast information, effectively assessing the structural and perceptual quality of the generated images. The SSIM between $\mathbf{R}$ and $\mathbf{F}$ is expressed as follows \cite{1284395}:
\begin{equation}
    \mathrm{SSIM} = \frac{(2\mu_{R}\mu_{F}+C_1)(2\sigma_{R,F}+C_2)}{(\mu_{R}^{2}+\mu_{F}^{2}+C_1)(\sigma_{R}^{2}+\sigma_{F}^{2}+C_2)},
\end{equation}
where $\mu_{R}$ and $\mu_{F}$ respectively denote the mean values of $\mathbf{R}$ and $\mathbf{F}$, $\sigma_{R}^{2}$ and $\sigma_{F}^{2}$ respectively denote the variance values of $R$ and $F$, and $\sigma_{R,F}$ represents the covariance variance values between $R$ and $F$. To avoid the denominator to be zero, constant parameters $C_1$ and $C_2$ are applied. In this work, $C_1 = 1 \times 10^{-4}$ and $C_2 = 9 \times 10^{-4}$ {\color{black}\cite{1284395}. For clarity, a lower NMSE value indicates a smaller error between the predicted and ground-truth images, whereas a higher SSIM value (closer to 1) reflects greater structural similarity. Together, an NMSE near 0 and an SSIM near 1 signify high-quality image reconstructions.}

\subsection{Numerical Evaluation of the Network Performance}
\label{sec:Numerical Evaluation}

To evaluate the performance of the proposed generator, 10 testing sample sets are randomly selected from the testing dataset. The results are as shown in Fig. \ref{fig:numericResult1}. Figs. \ref{fig:numericResult1}(a) and \ref{fig:numericResult1}(b) present the obstructed targets and the image reconstructions obtained using the conventional method outlined in (\ref{eq2_ls}). Similarly, Fig. \ref{fig:numericResult1}(c) illustrates the clean targets corresponding to the targets of interest in Fig. \ref{fig:numericResult1}(a), and their image reconstructions retrieved employing the conventional method are shown in Fig. \ref{fig:numericResult1}(d). Fig. \ref{fig:numericResult1}(e) shows the image reconstructions generated by the proposed cGAN-STM using the back-scattered measurements of the obstructed targets depicted in Fig. \ref{fig:numericResult1}(a). The results demonstrate that cGAN-STM successfully uses the back-scattered measurements of the obstructed targets to generate the image reconstructions of the targets of interest. Comparing Figs. \ref{fig:numericResult1}(d) and \ref{fig:numericResult1}(e), it is possible to observe a close agreement between the image reconstructions obtained by cGAN-STM from the back-scattered measurement from obstructed targets, and the reconstructions of unobstructed targets using the conventional technique. Furthermore, as previously discussed, the image reconstructions in Fig. \ref{fig:numericResult1}(b) are based on back-scattered measurements from the obstructed targets but obtained by the conventional method. Fig. \ref{fig:numericResult1}(b) shows that the conventional method yields poor quality reconstructions when applied to obstructed targets. In particular, when the targets are obstructed, they cannot be accurately identified from the reconstructions obtained by the conventional method. In contrast, employing the proposed cGAN-STM, the quality of the generated images (Fig. \ref{fig:numericResult1}(e)) improves significantly, producing results comparable to those obtained from unobstructed measurements.

With the testing dataset, the average NMSE and SSIM are respectively calculated to be 0.066 and 0.876. The proposed cGAN-STM achieves an average reconstruction time of 0.0314 seconds on an Intel\textsuperscript{\textregistered} Core\textsuperscript{TM} i7-1265U CPU, which indicates that the computational cost is moderate and suitable for real-time applications. More importantly, cGAN-STM can reconstruct targets under obstruction, which conventional methods are fundamentally unable to accomplish.

Additionally, to provide a complementary evaluation of reconstruction fidelity, a secondary analysis is conducted by directly comparing the reconstructed outputs from the testing dataset with the unobstructed and obstructed targets. Specifically, the NMSE and SSIM values were calculated for three cases:
\begin{enumerate}
    \item Reconstructions obtained using the conventional method with back-scattered measurements of obstructed targets.
    \item Reconstructions obtained using the conventional method with back-scattered measurements of unobstructed targets.
    \item Reconstructions generated by the proposed cGAN-STM using back-scattered measurements of obstructed targets.
\end{enumerate}

The quantitative results are summarized in Table \ref{tab:secana}. It is important to note that during training, the cGAN-STM learns a mapping from back-scattered echo measurements to the reconstruction space defined by the conventional imaging algorithm, rather than directly to the true target image. Consequently, the theoretical upper bound of the reconstruction quality is inherently constrained by the fidelity of the conventional method itself. The purpose of our approach is therefore not to surpass the conventional method but to compensate for the degradation and information loss that occur under obstructed conditions. From this perspective, achieving an SSIM of 0.543 and an NMSE of 0.226 under occlusions suggests a meaningful recovery relative to the conventional method.

\begin{table}[ht]
\caption{Comparison of Metric Values between Reconstructed and Target Images Using the Proposed and Conventional Methods}
\centering
\setlength{\tabcolsep}{15.4pt}
\begin{tabular*}{\columnwidth}{|c|c|c|c|}
\hline
\multicolumn{1}{|c|}{\textbf{Method}} &\multicolumn{1}{|c|}{\textbf{\begin{tabular}[c]{@{}c@{}}cGAN-STM\\ (proposed)\end{tabular}}}&\multicolumn{2}{|c|}{\textbf{Conventional Method}} \\ \hline
\multicolumn{1}{|c|}{\textbf{Obstruction}} &\multicolumn{2}{|c|}{\textbf{With}}&\multicolumn{1}{|c|}{\textbf{Without}} \\ \hline
\textbf{NMSE} & \textbf{0.543} & 0.454 & 0.635\\ \hline
\textbf{SSIM} & \textbf{0.226} & 0.345 & 0.158\\ \hline
\end{tabular*}
\label{tab:secana}
\end{table}

These findings demonstrate that by learning the features of the back-scattered measurements of the obstructed targets, the proposed generator is capable of removing the obstructing objects that occlude the targets of interest as well as generating high-fidelity image reconstructions. 

\subsection{Experimental Evaluation of the Network Performance}
\label{sec:Experimental Evaluation}
\begin{figure}[ht]
\centering
    \includegraphics[width=\columnwidth]{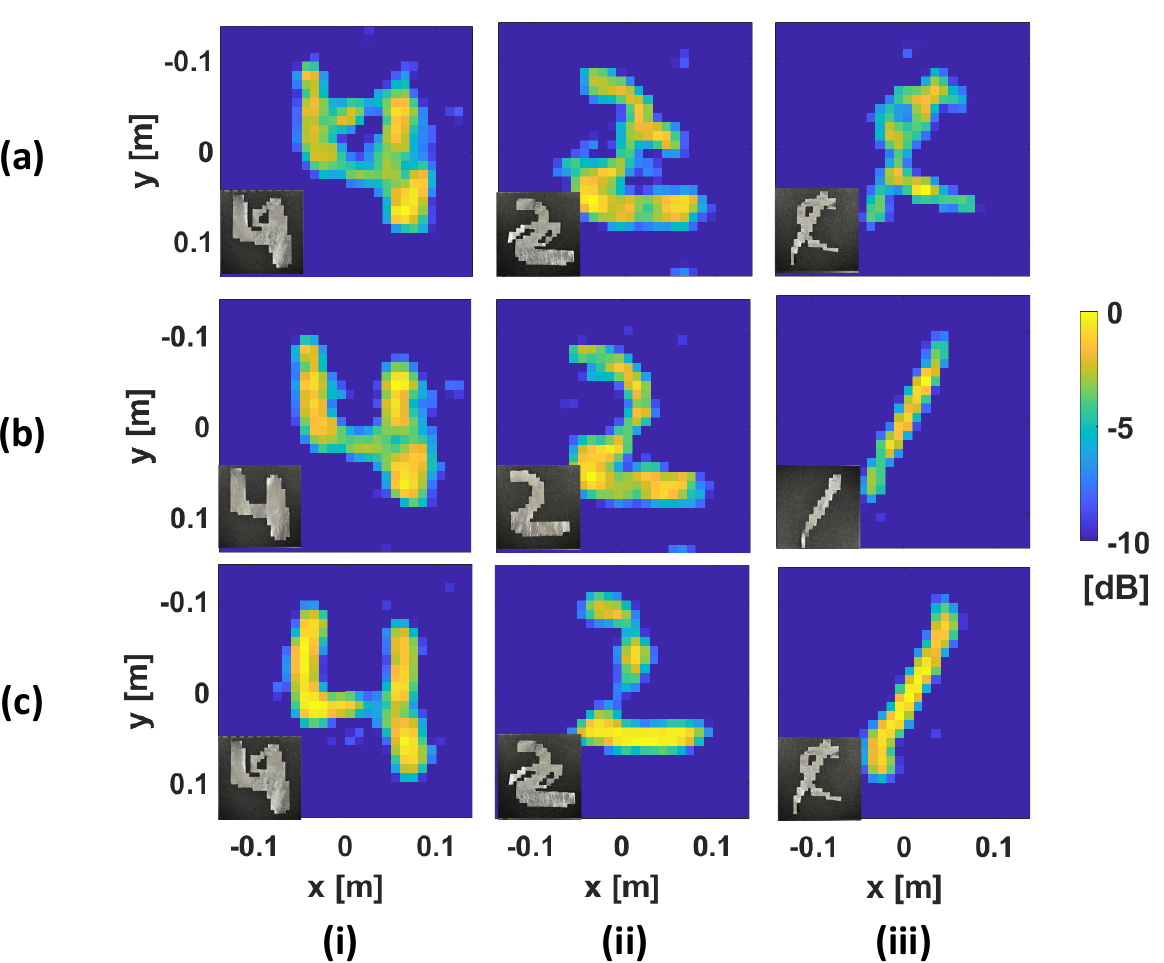} 
    \caption{Comparison of the reconstructed images with both experimentally measured sensing matrix and back-scattered measurements. (a) and (b) are generated by the conventional method outlined in (\ref{eq2_ls}), (c) is generated by cGAN-STM. The imaging targets are displayed in the bottom left corner. {\color{black}The colour scale represents normalized reflectivity in dB. The maximum (1) is displayed at $0$ dB, and values below $-10$ dB are clipped.}
    }
     \label{fig:ev}
\end{figure}

\begin{figure}[ht]
\centering
    \includegraphics[width=\columnwidth]{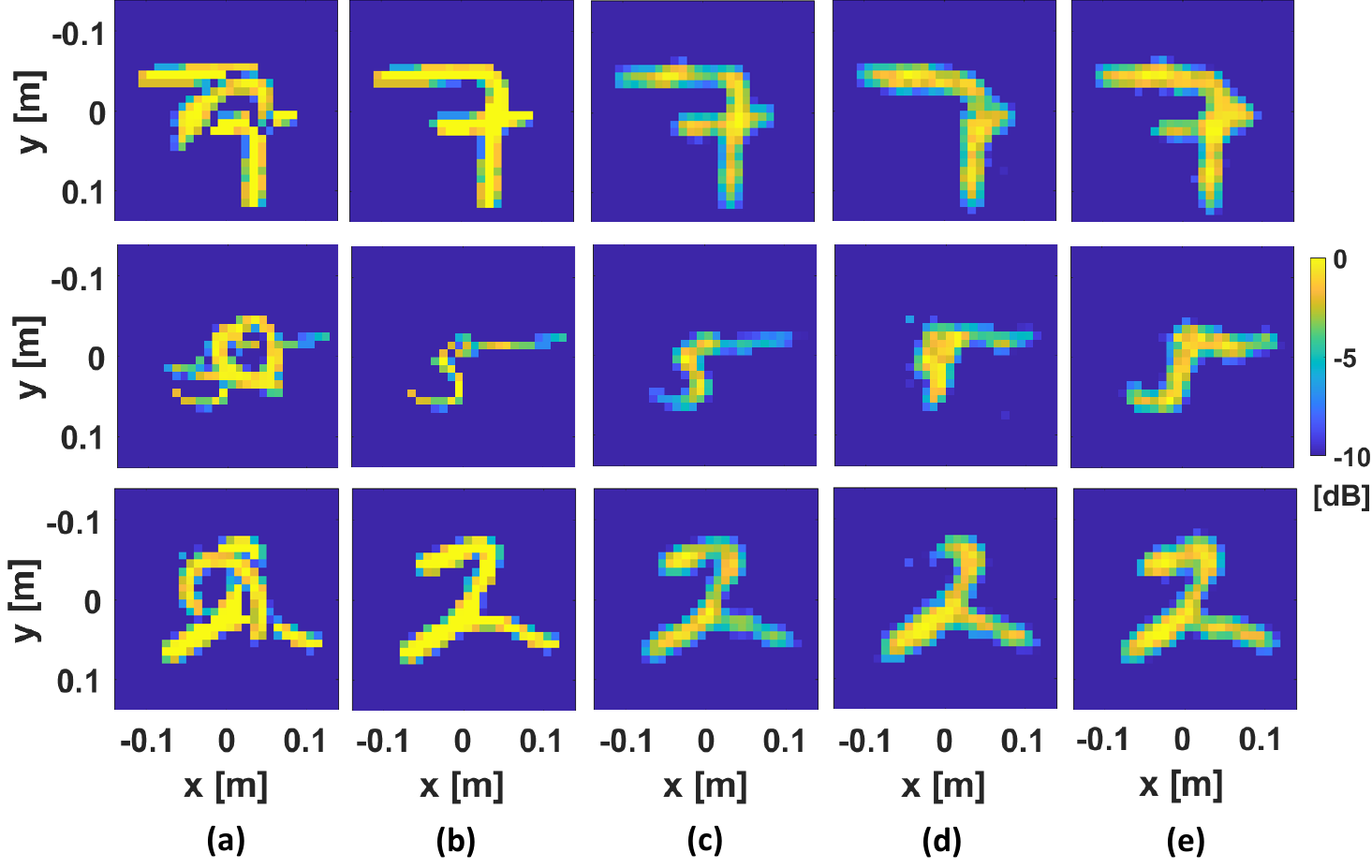} 
    \caption{Comparison of the reconstructed images based on cGAN-STM and cGAN-STM-Lite: (a) obstructed targets; (b) targets of interest (i.e., unobstructed targets); (c) image reconstructions of the targets of interest using the conventional method (4); and image reconstructions of the obstructed targets using (d) cGAN-STM-Lite and (e) cGAN-STM. {\color{black}The colour scale represents normalized reflectivity in dB. The maximum (1) is displayed at $0$ dB, and values below $-10$ dB are clipped.}
    }
     \label{fig:ab}
\end{figure}

\begin{figure*}[htb]
\centering
    \includegraphics[width=\textwidth]{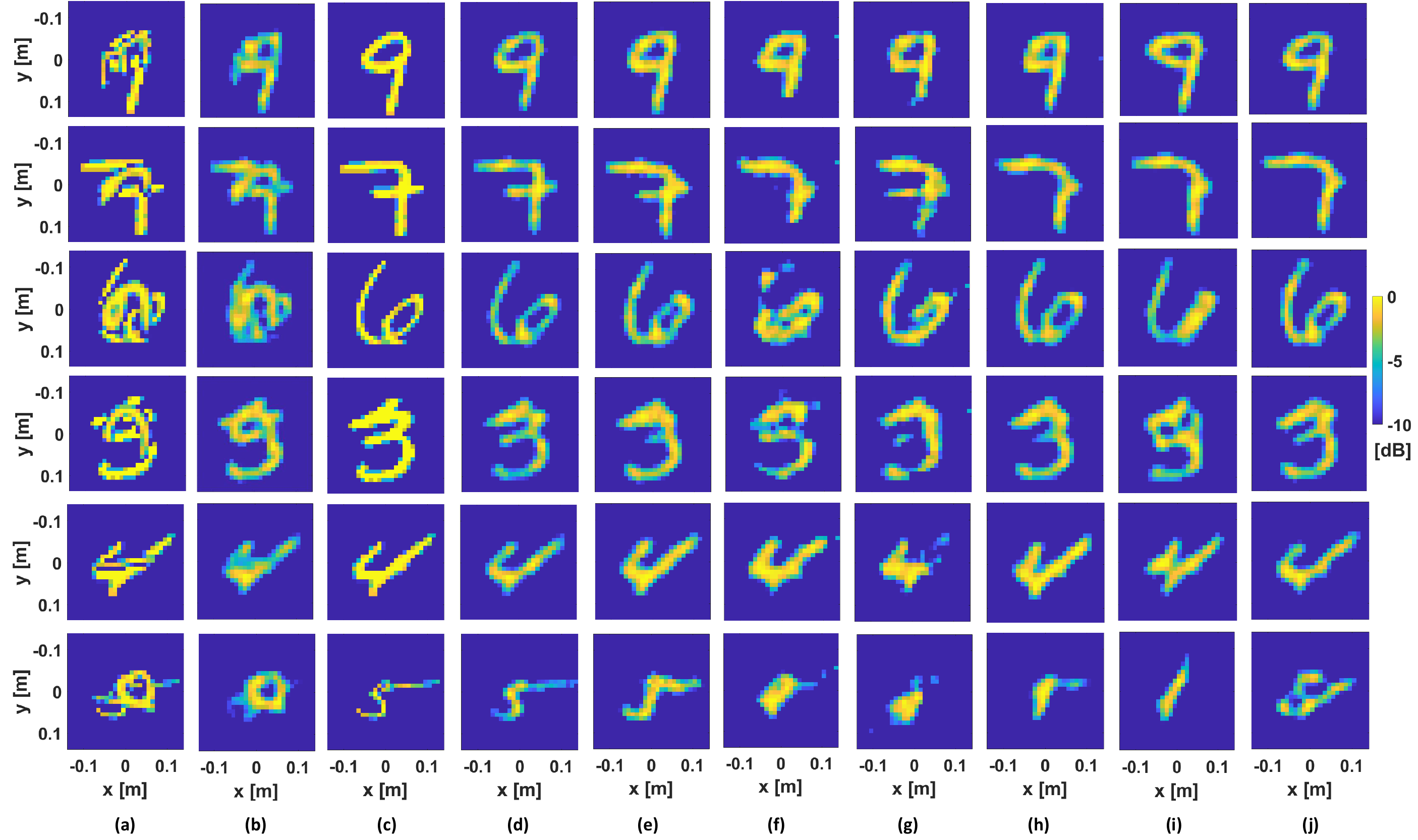} 
    \caption{Comparison of the reconstructed images: (a) obstructed targets; (b) the reconstructed images for the obstructed targets, generated by the conventional method (\ref{eq2_ls}); (c) targets of interest; (d) the reconstructed images for the targets of interest, generated by the conventional method (\ref{eq2_ls}); (e), (f), (g), (h), (i), (j) reconstructed images of the targets of interest from (a), generated by cGAN-STM (proposed), Att-GAN \cite{9676480}, DCCGAN \cite{10473754}, ASR-T \cite{10056769} and U-Net \cite{10633872}{\color{black}, and V-Net \cite{11030585}}, respectively. {\color{black}The colour scale represents normalized reflectivity in dB. The maximum (1) is displayed at $0$ dB, and values below $-10$ dB are clipped.}}
     \label{fig:Method_Comparison}
\end{figure*}

To further evaluate the performance of cGAN-STM in practical scenarios, the imaging setup depicted in Fig. \ref{fig:antenna} is employed. Back-scattered measurements of several obstructed targets are acquired. The image reconstructions of these obstructed targets obtained using the conventional method in (\ref{eq2_ls}) are presented in Fig. \ref{fig:ev}(a). The images of the corresponding targets of interest, without obstructions, obtained also with the conventional method are shown in Fig. \ref{fig:ev}(b). Fig. \ref{fig:ev}(c) illustrates the reconstructions of the obstructed targets obtained by the proposed cGAN-STM. The corresponding imaging targets are indicated in the bottom-left corner of each reconstruction. NMSE and SSIM values are calculated by comparing Figs. \ref{fig:ev}(b) and \ref{fig:ev}(c) and are summarized in Table \ref{tab:ev}. These results demonstrate the feasibility of applying cGAN-STM in practical settings.

\begin{table}[h]
\caption{Quantitative comparison for the results presented in Fig. \ref{fig:ev}.}
\centering
\setlength{\tabcolsep}{20.13pt}
\begin{tabular*}{\columnwidth}{|c|c|c|c|}
\hline
\textbf{Target No.} & (i) & (ii) & (iii) \\ \hline
\textbf{NMSE} & 0.221  & 0.273  &  0.235 \\ \hline
\textbf{SSIM} &  0.610 &  0.522 &  0.520 \\ \hline
\end{tabular*}
\label{tab:ev}
\end{table}

\subsection{Ablation Study}
\label{sec:ab}

To evaluate the impact of the STMs, an ablation study is conducted. A baseline model, termed cGAN-STM-Lite, is constructed with the same architecture as cGAN-STM but without the STMs, and included for comparison. Fig. \ref{fig:ab} summarizes the reconstruction results. Specifically, Figs. \ref{fig:ab}(a) and \ref{fig:ab}(b) show the obscured targets and their corresponding unobstructed counterparts, respectively. Fig. \ref{fig:ab}(c) presents the reconstruction obtained by the conventional method in (\ref{eq2_ls}) using the unobstructed targets in Fig. \ref{fig:ab}(b). In contrast, Figs. \ref{fig:ab}(d) and \ref{fig:ab}(e) display the reconstructions obtained by cGAN-STM-Lite and cGAN-STM, respectively, using the obscured targets in Fig. \ref{fig:ab}(a).

Although the cGAN-STM-Lite results in Fig. \ref{fig:ab}(d) capture the target outlines, visual comparison shows that the corresponding cGAN-STM results in Fig. \ref{fig:ab}(e) provide better resolution and visual quality. Furthermore, the average NMSE and SSIM values for cGAN-STM-Lite are listed in Table \ref{tab:ab}. The results presented in Fig. \ref{fig:ab} (qualitative) and Table II (quantitative) between cGAN-STM-Lite and cGAN-STM confirm that the inclusion of STMs improves the reconstruction quality, highlighting their effectiveness in information filtering.

\begin{table}[ht]
\caption{Comparison of metric values when with and without STMs}
\centering
\setlength{\tabcolsep}{20.6pt}
\begin{tabular*}{\columnwidth}{|c|c|c|}
\hline
\textbf{Network} & \textbf{\begin{tabular}[c]{@{}c@{}}cGAN-STM\\ (proposed)\end{tabular}} & \textbf{cGAN-STM-Lite} \\ \hline
\textbf{NMSE} & \textbf{0.066} & 0.084 \\ \hline
\textbf{SSIM} & \textbf{0.876} & 0.843 \\ \hline
\end{tabular*}
\label{tab:ab}
\end{table}

\subsection{Method Benchmarking}
\label{sec:Method Benchmarking}
As discussed in Section \ref{sec: Introduction}, a variety of deep learning techniques have been applied to address the image reconstruction problem in microwave imaging, as well as in CMI. To ensure a fair evaluation, cGAN-STM is benchmarked against representative state-of-the-art methods. Since the architecture of cGAN-STM is based on cGAN, two advanced cGAN architectures, Att-GAN \cite{9676480} and DCCGAN \cite{10473754}, are selected. A transformer-based approach proposed by \cite{10056769}, termed ASR-T, is also considered. In addition, given that the architecture of the proposed generator follows a U-Net architecture, a state-of-the-art U-Net variant \cite{10633872} and {\color{black}a state-of-the-art V-Net variant \cite{11030585} are adopted for comparison, as both were proposed for image reconstruction tasks.} Other approaches, such as diffusion-based models \cite{ho2020denoising}, are not considered because of their reliance on hundreds of iterative denoising steps, which makes them computationally too expensive for CMI applications, where rapid or near real-time reconstruction is typically required.

The average values of NMSE and SSIM are given in Table \ref{tab: methods}. As shown in Table \ref{tab: methods}, the NMSE value provided by cGAN-STM is lower than the values achieved when the other four approaches are employed. Similarly, the SSIM value calculated using the proposed cGAN-STM is higher than those provided by the other four methods, suggesting a better performance on the imaging problem involving obstructed imaging targets.

\begin{table}[h]
\caption{Comparison of metric values between different methods}
\centering
\setlength{\tabcolsep}{3.5pt}
\begin{tabular*}{\columnwidth}{|c|c|c|c|c|c|c|}
\hline
\textbf{Network} & \textbf{\begin{tabular}[c]{@{}c@{}}cGAN-STM\\ (proposed)\end{tabular}}
& \textbf{\begin{tabular}[c]{@{}c@{}}Att-GAN\\ \cite{9676480}\end{tabular}}
& \textbf{\begin{tabular}[c]{@{}c@{}}DCCGAN\\ \cite{10473754}\end{tabular}} 
& \textbf{\begin{tabular}[c]{@{}c@{}}ASR-T\\ \cite{10056769}\end{tabular}}
& \textbf{\begin{tabular}[c]{@{}c@{}}U-Net\\ \cite{10633872}\end{tabular}}
& \textbf{\begin{tabular}[c]{@{}c@{}}{\color{black}V-Net}\\ {\color{black}\cite{11030585}}\end{tabular}}
 \\ \hline
\textbf{NMSE} & \textbf{0.066}  & 0.244  & 0.163 & 0.074 & 0.099 & {\color{black}0.082 }\\ \hline
\textbf{SSIM} & \textbf{0.876}  &  0.747 &  0.721 & 0.854 & 0.810 & {\color{black}0.850 }\\ \hline
\end{tabular*}
\label{tab: methods}
\end{table}

To further compare the performance of different approaches, several testing sample sets are selected and presented in Fig. \ref{fig:Method_Comparison}. Fig. \ref{fig:Method_Comparison}(a) shows the obstructed imaging targets, while Fig. \ref{fig:Method_Comparison}(c) shows the targets of interest (unobstructed) corresponding to the obstructed imaging targets in Fig. \ref{fig:Method_Comparison}(a). Figs. \ref{fig:Method_Comparison}(b) and \ref{fig:Method_Comparison}(d) respectively present the image reconstructions of the imaging targets shown in Figs. \ref{fig:Method_Comparison}(a) and \ref{fig:Method_Comparison}(c), generated by the conventional method. As can be observed, when the targets of interest are not obstructed, the conventional approach provides high-fidelity reconstructions of the targets (Fig. \ref{fig:Method_Comparison}(d)). In contrast, when the targets of interest are obstructed, the targets of interest cannot be accurately imaged (Fig. \ref{fig:Method_Comparison}(b)). Similarly, Figs. \ref{fig:Method_Comparison}(e), \ref{fig:Method_Comparison}(f), \ref{fig:Method_Comparison}(g), \ref{fig:Method_Comparison}(h),
\ref{fig:Method_Comparison}(i){\color{black}, and \ref{fig:Method_Comparison}(j)} depict the image reconstructions using the back-scattered measurements from the obstructed imaging targets shown in Fig. \ref{fig:Method_Comparison}(a), generated by cGAN-STM (proposed), Att-GAN \cite{9676480}, DCCGAN \cite{10473754}, ASR-T \cite{10056769}, U-Net \cite{10633872}{\color{black}, and V-Net \cite{11030585}}, respectively. It can be observed from Fig. \ref{fig:Method_Comparison} that the proposed cGAN-STM achieves higher image quality in reconstructing certain targets, such as the numbers 5, 6 and 7, compared to the other five methods. These qualitative (Fig. \ref{fig:Method_Comparison}) and quantitative (Table \ref{tab: methods}) results demonstrate that the cGAN-STM is capable of providing more accurate image reconstructions when the imaging objects are obstructed.

\subsection{Obstruction-free Scenario Analysis}
\label{sec:Obstruction size1}
\begin{figure}[tb]
\centering
    \includegraphics[width=\columnwidth]{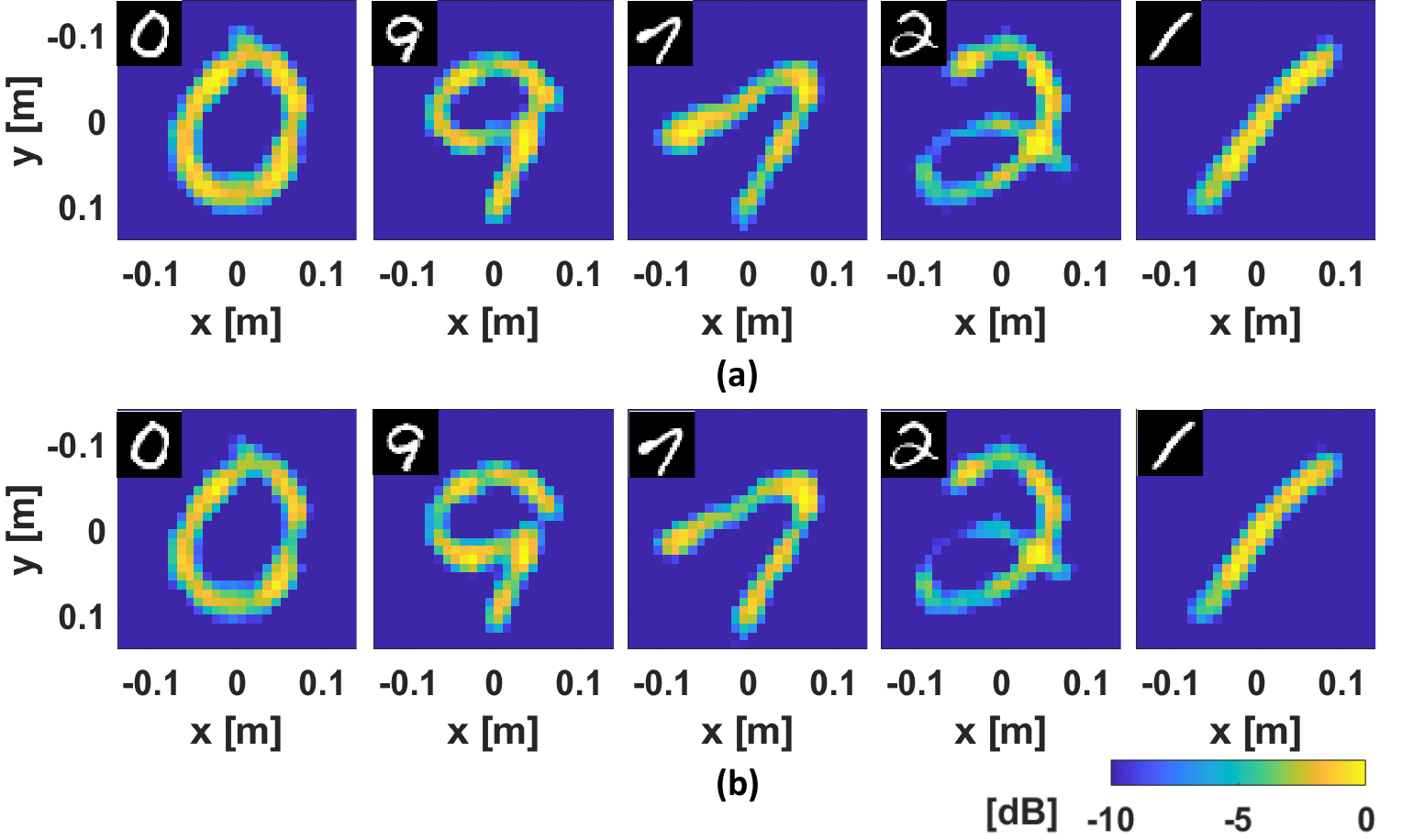} 
    \caption{Comparison of image reconstructions using the experimentally measured sensing matrix and numerically synthesized back-scattered measurements of unobstructed targets, generated from: (a) conventional method (outlined in (\ref{eq2_ls})) and (b) the proposed generator. The unobstructed imaging targets are displayed in the upper left corner. The colour scale represents normalized reflectivity in dB. The maximum (1) is displayed at $0$ dB, and values below $-10$ dB are clipped.}
     \label{fig:singletest}
\end{figure}

As discussed above, the imaging targets of interest are the digits and the obstructing objects are the letters. Thus, to validate the information learnt by cGAN-STM, the testing dataset containing the reconstructed images of the unobstructed targets of interest and their corresponding back-scattered measurements are employed. Additionally, a comparison of the image reconstructions using back-scattered measurements of the unobstructed targets of interest generated by the conventional methods and those produced by cGAN-STM is shown in Fig. \ref{fig:singletest}. Figs. \ref{fig:singletest}(a) and \ref{fig:singletest}(b) respectively illustrate the image reconstructions by the conventional methods and the proposed cGAN-STM with the back-scattered measurements of the targets of interest. The average NMSE and SSIM values of 10000 testing sample sets associated with sole targets of interest are 0.033 and 0.927, respectively. Notably, for this testing dataset, the average NMSE value is lower and the average SSIM value is higher when compared to the results for obstructed targets. These results demonstrate that cGAN-STM effectively retrieves the information relevant to the targets of interest, regardless of the presence or absence of obstruction.

\subsection{Impact of Different Size of Obstruction}
\label{sec:size}
\begin{figure}[!t]
    \centering
    \includegraphics[width=\columnwidth]{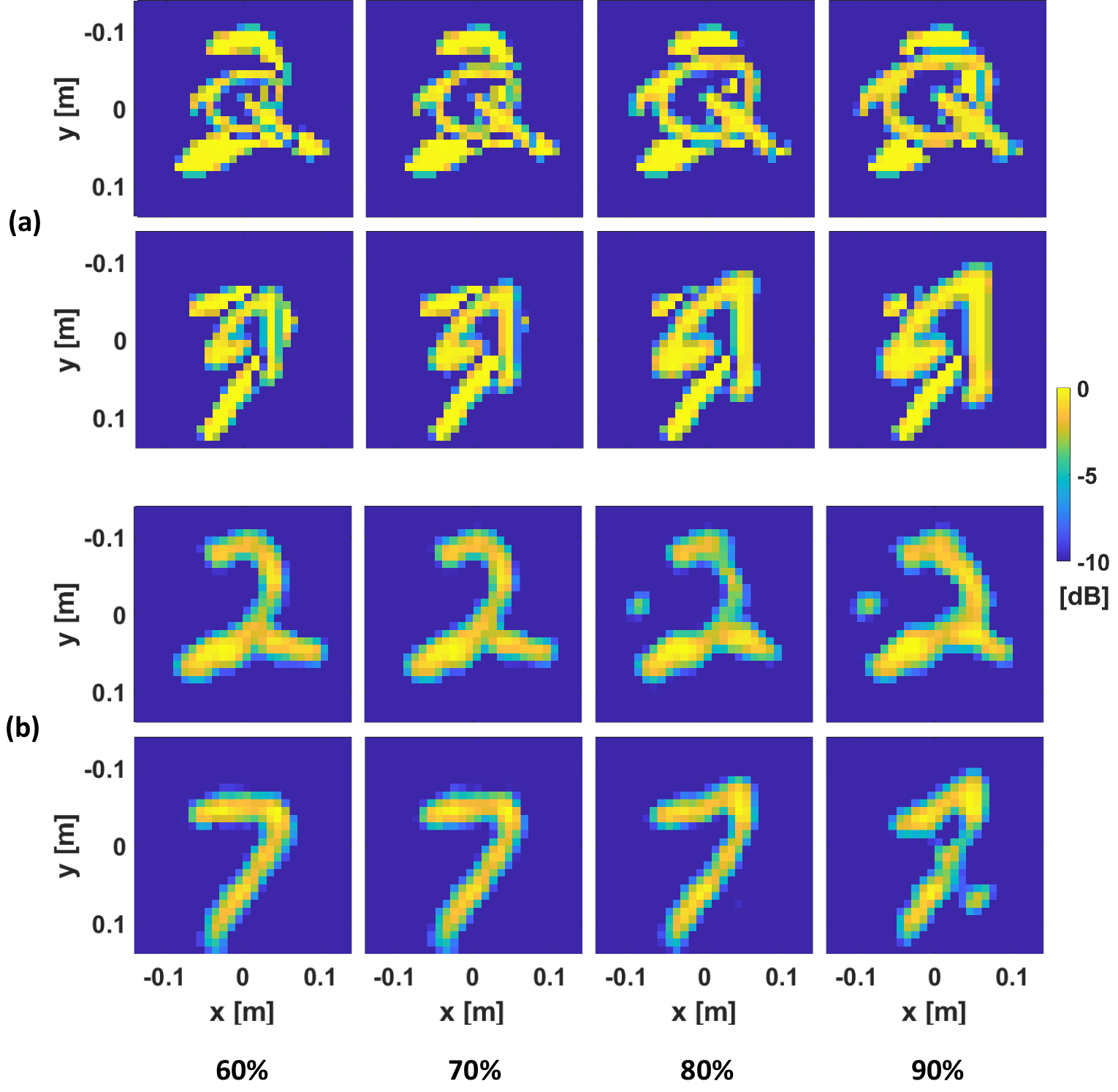}
    \caption{Comparison of cGAN-STM's performance with different sizes of obstruction: (a) obstructed targets; (b) image reconstructions generated by cGAN-STM. The colour scale represents normalized reflectivity in dB. The maximum (1) is displayed at $0$ dB, and values below $-10$ dB are clipped.}
    \label{fig:scales}
\end{figure}

As previously discussed, for dataset generation, the size of obstructions is set to 60\% of that of the targets of interest. In this section, the impact of varying obstruction sizes on the cGAN-STM's performance is evaluated. The sizes of obstructions range from 60\% to 90\% of the targets of interest size, with each size corresponding to 10000 samples. The average values of NMSE and SSIM calculated for different sizes of obstructions are given in Table \ref{tab:scales}. The increase in NMSE and decrease in SSIM indicate a decline in the quality of image reconstructions generated by cGAN-STM as the size of the obstructions increases. To visualize this, as shown in Fig. \ref{fig:scales}, two obstructed targets are randomly chosen, where the size of the obstruction is changed accordingly. Fig. \ref{fig:scales}(a) shows the obstructed targets where the obstruction sizes are different. Fig. \ref{fig:scales}(b) illustrates the corresponding image reconstructions generated by the proposed cGAN-STM. 

\begin{table}[h]
\caption{Comparison of metric values between different sizes of obstructions}
\centering
\setlength{\tabcolsep}{13.9pt}
\begin{tabular*}{\columnwidth}{|c|c|c|c|c|}
\hline
\textbf{\begin{tabular}[c]{@{}c@{}}Obstruction\\ Ratio\end{tabular}} & \textbf{60\%} & \textbf{70\%} & \textbf{80\%} & \textbf{90\%}\\ \hline
\textbf{NMSE} & 0.066  & 0.122  & 0.385  & 0.540\\ \hline
\textbf{SSIM} & 0.876  & 0.810  & 0.565  & 0.473\\ \hline
\end{tabular*}
\label{tab:scales}
\end{table}

The results demonstrate that with the increase in size of obstructions, an increased amount of the key features of the targets of interest are concealed. In this situation, the received back-scattered measurements lack crucial target information from the occluded regions. As a result, cGAN-STM struggles to learn the useful information associated with the missing key features, logically leading to a degradation in the quality of reconstructions of the targets of interest. However, as can be observed in Fig. \ref{fig:scales}, even when the size of the obstructing targets is 70\% of that of the targets of interest, the quality of the reconstructed images is high.

\par
\normalcolor
\subsection{Analysis of STM Mechanism}
\label{Sec:Analysis of STM Mechanism}
\color{black}
To further analyze the internal behavior of STM, both quantitative and qualitative analyses are conducted:
\subsubsection{Visualization and Resolution} 
\begin{figure}
    \centering
    \includegraphics[width=\linewidth]{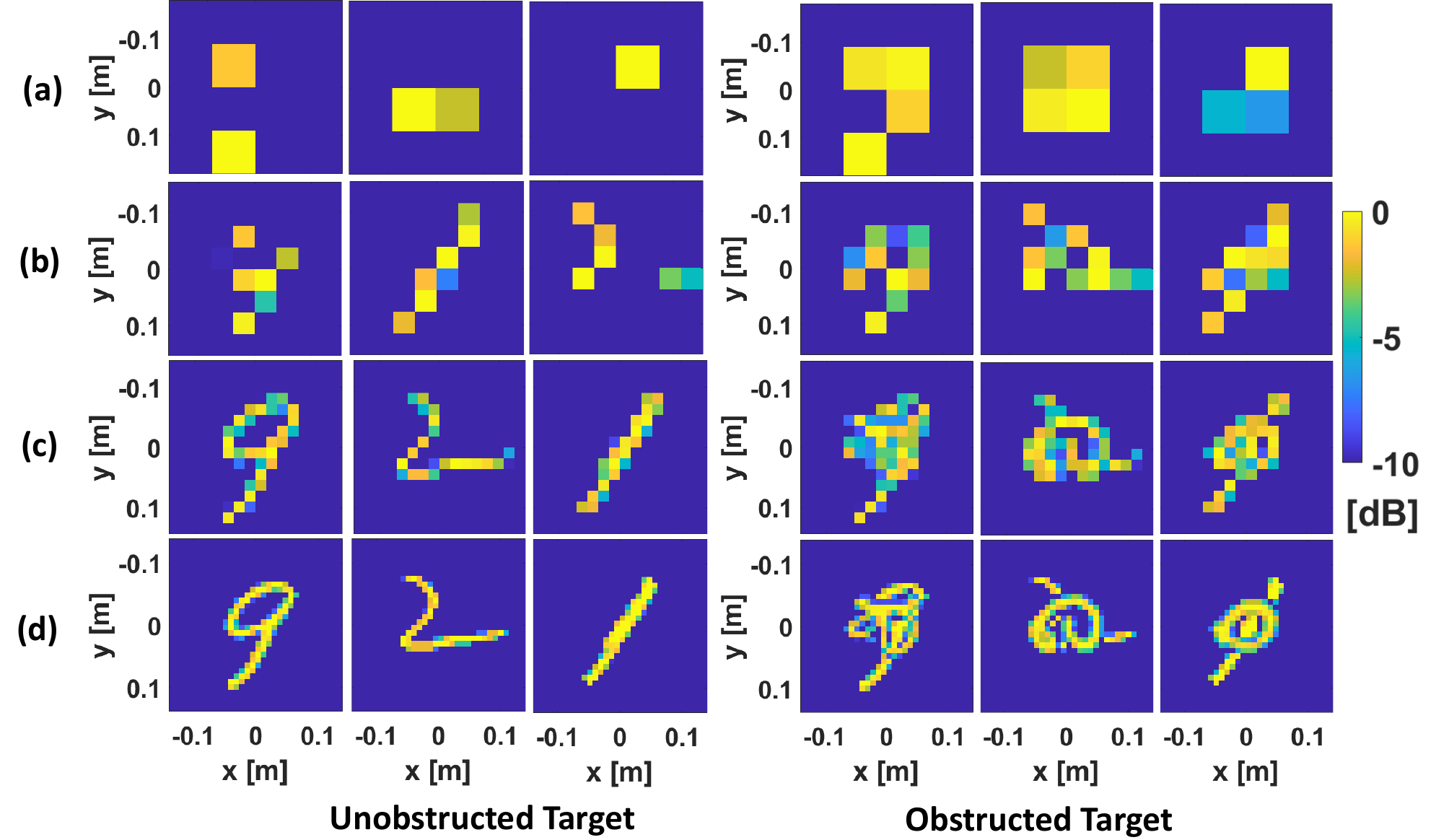}
    \caption{Resized targets with the size of (a) 4 $\times$ 4; (b) 8 $\times$ 8; (c) 16 $\times$ 16; (d) 32 $\times$ 32.}
    \label{fig:tgt_resize}
\end{figure}

Since STM modules are connected to the decoder, the spatial resolution of feature maps increases progressively from 2$\times$2 to 32$\times$32, while the target size is 28$\times$28. To enable consistent visual comparison, target images are resized accordingly. As shown in Fig. \ref{fig:tgt_resize}, when the feature maps are smaller than 8$\times$8, the targets become visually indistinguishable due to excessive down-sampling. Therefore, the subsequent analyses focus on STM outputs with 16$\times$16 and 32$\times$32 resolutions, where both target and obstruction regions remain distinguishable for statistical evaluation.

\subsubsection{Suppression Score Definition} To evaluate the suppression effect of STM at the two representative resolutions, a suppression score $S_{m}$ is defined to quantify the relative reduction in activation magnitude within a given semantic region and feature map:
\begin{equation}
    S_m = \frac{1}{T}\sum^{T}_{i=1}\frac{(|F_{in}|(x_i, y_i) - |F_{out}|(x_i, y_i))}{|F_{in}|(x_i, y_i) + \varepsilon}\odot m(x_i, y_i),
\end{equation}
where $(x_i,y_i)$ denotes the pixel coordinate, $|F_{in} |$ and $|F_{out}|$ denote the magnitude of the input and output feature maps, respectively. A small constant $\varepsilon$ = $10^{-8}$ is added to the denominator to avoid numerical instability when the normalization factor approaches zero. The binary mask $m$ is derived from annotations that label obstruction and target-visible regions, ensuring a clear semantic separation for quantitative comparison. $T$ denotes the total number of valid pixels within the mask. Normalizing the difference between $|F_{in} |$ and $|F_{out}|$  converts it into a relative change with respect to the input magnitude, thereby mitigating the influence of individual sample amplitudes on cross-sample comparisons. The magnitude values are applied because the soft-thresholding mechanism acts on the magnitude of activations rather than their sign. The higher value of $S_m$ indicates a stronger suppression. 

To assess adaptive selectivity, the per-feature-map suppression difference between obstruction and target-visible regions is defined as:
\begin{equation}
    \Delta S = S_{m_{ob}} - S_{m_{tg}},
\end{equation}
where $m_{ob}$ and $m_{tg}$ correspond to suppression scores in obstruction and target-visible areas, respectively.
\begin{figure}
    \centering
    \includegraphics[width=\linewidth]{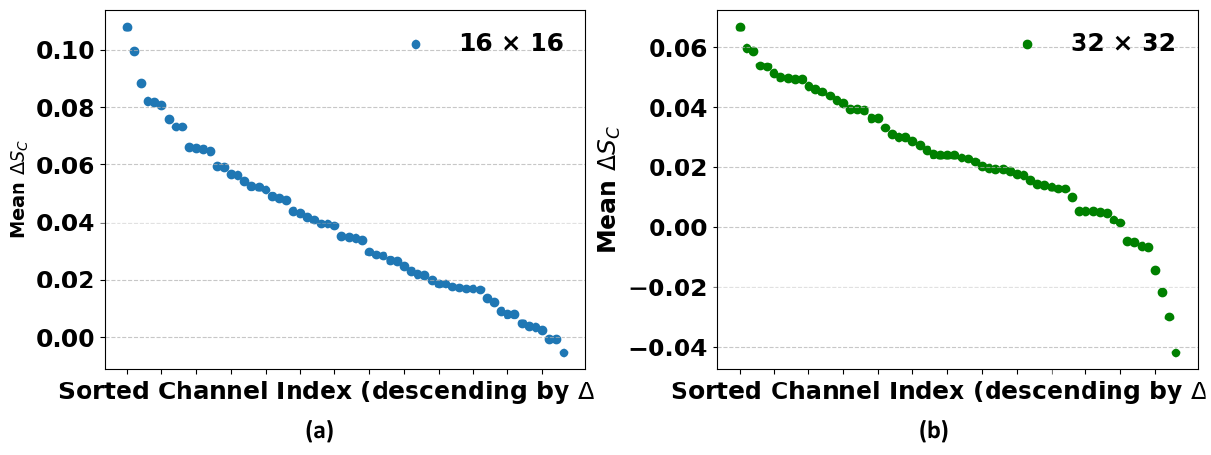}
    \caption{The average feature-map-wise mean $\Delta S$ across all samples sorted in descending order, with different size of feature maps: (a) 16 $\times$ 16; and (b) 32 $\times$ 32.}
    \label{fig:suppresiongap}
\end{figure}
\subsubsection{Quantitative Feature-map-wise Analysis}
To identify representative feature maps for visualization, a quantitative analysis of feature map-wise suppression behavior is conducted using the per-feature-map suppression difference $\Delta S$. For each STM size, $\Delta S$ is computed per feature map and averaged across samples, then sorted in descending order. The averaged $\Delta S$ values for two feature map sizes are shown in Fig. \ref{fig:suppresiongap}. As observed, $\Delta S$ exhibits a non-flat distribution across feature maps, indicating that the STM modulates feature suppression in a feature-map-dependent manner rather than performing uniform attenuation.

\subsubsection{Qualitative Feature Map Analysis}
\begin{figure*}
    \centering
    \includegraphics[width=\linewidth]{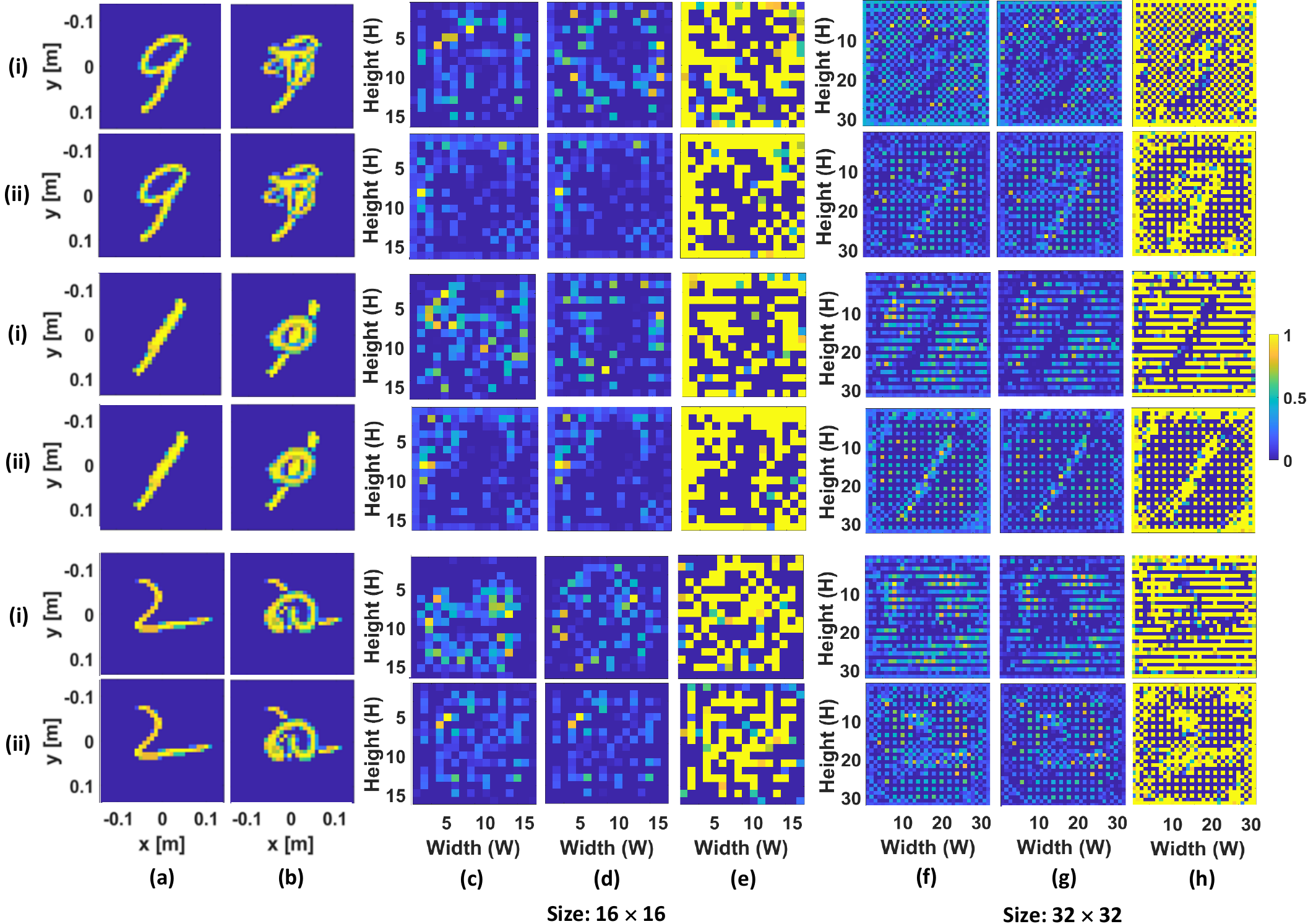}
    \caption{Comparison of (i) top-ranked feature map and (ii) bottom-ranked feature map with three example targets. For the $16 \times 16$ size, (c) input feature maps; (d) output feature maps; and (e) Difference Heat Map. For the $32 \times 32$ size, (f) input feature maps; (g) output feature maps; and (h) Difference Heat Map. 
    }
    \label{fig:Feature_Map}
\end{figure*}

Feature maps with positive $\Delta S$ correspond to stronger suppression in obstruction regions than in target-visible regions, whereas negative $\Delta S$ indicates stronger suppression in target-visible regions. To interpret these effects, representative feature maps from three test samples are visualized in Fig. \ref{fig:Feature_Map}. Figs. \ref{fig:Feature_Map}(a) and (b) present the targets of interest and their obstructed views. Figs. \ref{fig:Feature_Map}(c), (d) and (e) present the input and output feature maps together with their suppression heat, at the 16 × 16 resolution, while Figs. \ref{fig:Feature_Map}(f), (g) and (h) show the corresponding results at the 32 × 32 resolution. Figs. \ref{fig:Feature_Map}(i) and (ii) indicate the top-ranked feature map and bottom-ranked feature map.

As shown in Fig. \ref{fig:Feature_Map}(i), the suppression heat map aligns with non-target areas (including obstruction and background), indicating that STM selectively suppresses activations unrelated to the target. This spatial correspondence is consistent across samples, suggesting a stable non-target suppression pattern. On the other hand, the suppression heat map of the bottom-ranked feature map $\Delta S$ shows that the suppression area covers the target area. To understand the behaviour of the suppression on target area, a quantitative analysis is performed using 32×32 feature maps. In this analysis, not only the cosine similarity \cite{xia2015learning} is used, a suppression strength ($r_s$) between the input and output feature maps within the target region is also used:
\begin{equation}
    r_s = 1 - \frac{1}{\sum m_{tg}}\sum_{(x,y):m_{tg}(x,y) = 1} \frac{|F_{out}|}{|F_{in}| + \varepsilon}.
\end{equation}

Across three examples, the cosine similarity values are 0.987, 0.991, and 0.994, and suppression strength values are 0.224, 0.175, and 0.127. These results indicate that STM mainly reduces feature magnitude while largely preserving the feature direction within the target region. This behavior aligns with prior studies showing that magnitude control enhances training stability and mitigates over-fitting \cite{bengio2017deep, moradi2020survey,tian2022comprehensive}.

\subsubsection{Analysis of the Learned Threshold Values $\tau$}
\begin{figure}
    \centering
    \includegraphics[width=\linewidth]{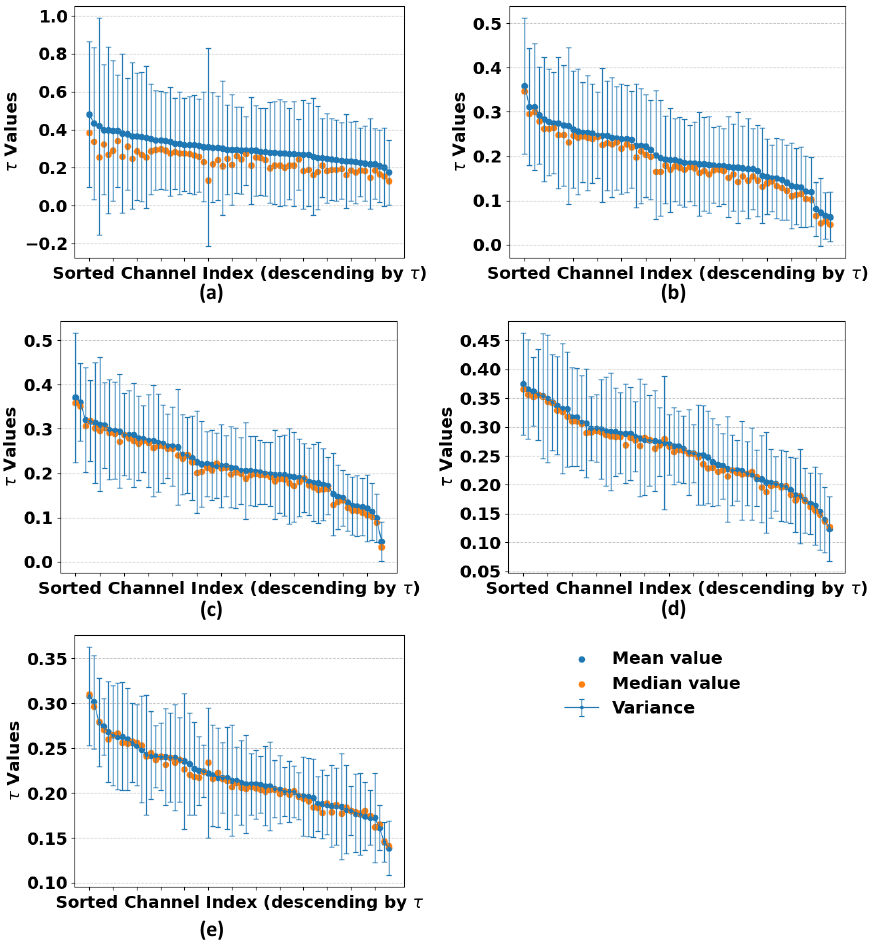}
    \caption{{\color{black}The average feature-map-wise mean $\tau$ across all samples at descending order, with different size of feature maps: (a) 2$\times$2; (b)4$\times$4; (c) 8$\times$8; (d) 16$\times$16 and (e) 32$\times$32.}}
    \label{fig:tau}
\end{figure}

\begin{figure}
    \centering
    \includegraphics[width=\linewidth]{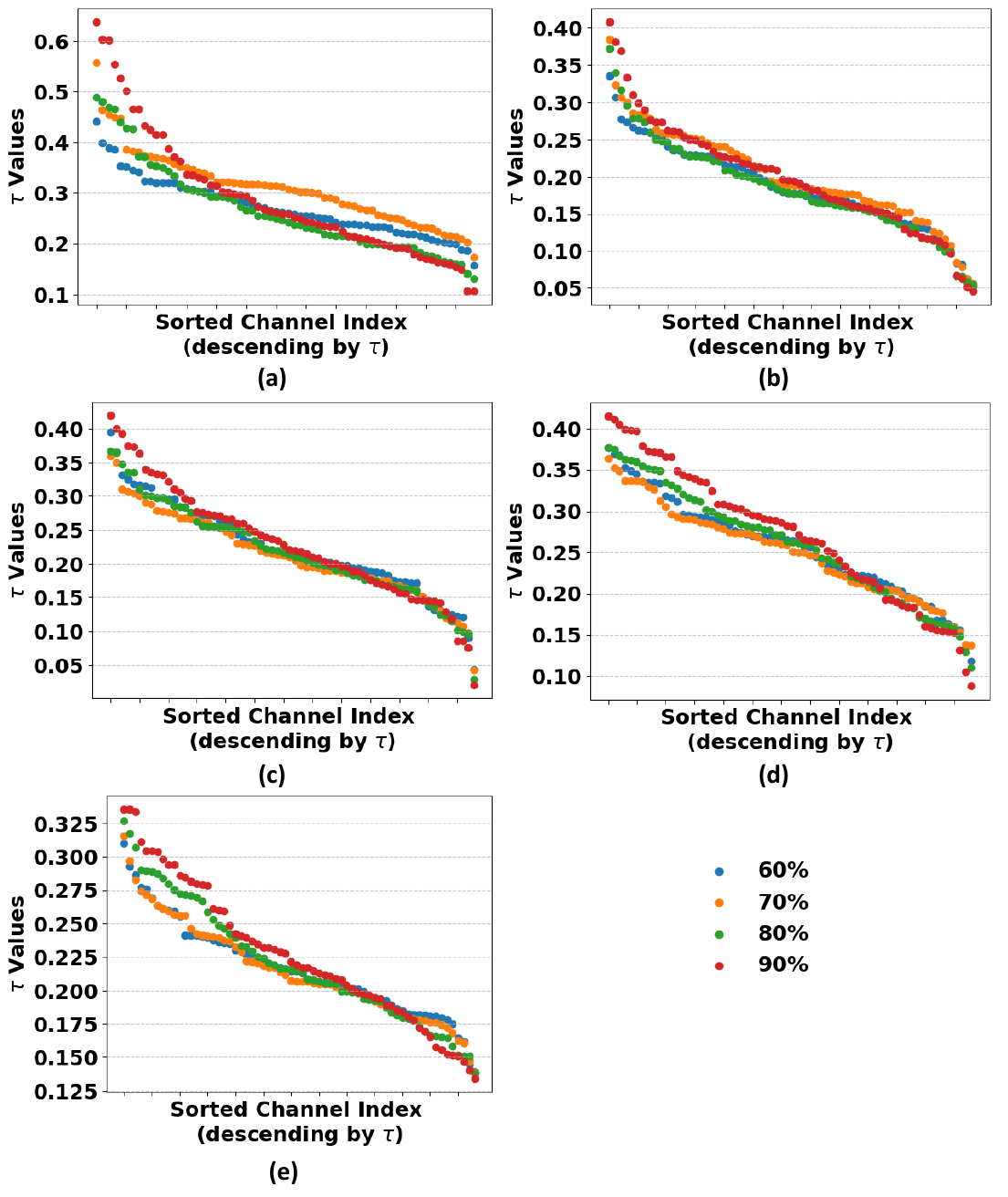}
    \caption{{\color{black}The average feature-map-wise mean $\tau$ across the samples with different size of obstructions at descending order, with different size of feature maps: (a) 2$\times$2; (b)4$\times$4; (c) 8$\times$8; (d) 16$\times$16 and (e) 32$\times$32.}}
    \label{fig:tau_diff_ob}
\end{figure}

As described in Section \ref{subsec:STMArchi}, the threshold value $\tau$ for each sample, layer, and feature map is generated by two fully connected networks, resulting in sample-dependent variations. To visualize this adaptability, the mean, median, and variance of $\tau$ across test samples are summarized in Fig. \ref{fig:tau}. The average feature-map-wise means, sorted in descending order for different feature map sizes, show substantial variations both across feature maps and samples. These results indicate that STM adapts its threshold values to input features rather than applying fixed or global parameters.
To further assess adaptability, obstruction types are characterized by varying obstruction sizes (60\%–90\% of the target area, as defined in Section \ref{sec:Obstruction size1}). As shown in Fig. \ref{fig:tau_diff_ob}, the mean $\tau$ per feature map changes with obstruction levels. Although the overall trend shows higher thresholds with increased obstruction, several layers display local non-monotonic variations. This behavior reflects STM’s sensitivity to local activation statistics rather than the obstruction ratio itself. When occlusions dominate spatially and become homogeneous, feature variance decreases, leading to smaller thresholds. Such non-monotonic adaptation confirms that STM acts as an input-dependent modulation mechanism rather than a static denoising operator.

To examine whether STM performs feature selection, the receiver operating characteristic–area under the curve (ROC–AUC) metric \cite{huang2005using} is computed. To begin with, a binary mask is defined by assigning 1 to non-target regions (including obstruction and background) and 0 to target-visible regions. By the way of this, an ROC-AUC value closer to 1 indicates a higher relationship between the suppression and the non-target region. The average ROC-AUC is calculated to be 0.701, indicating a consistent spatial bias toward non-target suppression. Given that STM operates without supervision on obstruction masks, this discrimination supports its role as an adaptive, semantically selective mechanism rather than a generic feature selector.

\subsubsection{Conclusion}
In summary, STM functions as an input-dependent suppression module that selectively attenuates non-target features without explicit obstruction recognition, thereby enhancing feature discrimination and model robustness.

\subsection{Noise Study}
\label{sec:Noise Study}
To assess the performance of the proposed cGAN-STM under noisy conditions, Gaussian white noise with SNR levels of 0, 5, 10, and 15 dB is added to the testing back-scattered measurements, due to its statistical similarity to typical radar noise \cite{5936732}. For each SNR level, 1000 randomly selected samples from the testing dataset are corrupted with Gaussian noise to generate the noisy inputs. Reconstruction performance is quantified using NMSE and SSIM, with results summarized in Table \ref{tab:noise}. It can be observed that the reconstruction quality decreases at lower SNR levels. At 15 dB, cGAN-STM achieves an NMSE of 0.168 and an SSIM of 0.700, indicating that the main structural features of the targets are largely preserved despite minor degradation relative to the noise-free case. At lower SNRs, such as 0 and 5 dB, quantitative performance deteriorates, showing that noise starts to affect the reconstruction quality.

\begin{figure}
    \centering
    \includegraphics[width=\columnwidth]{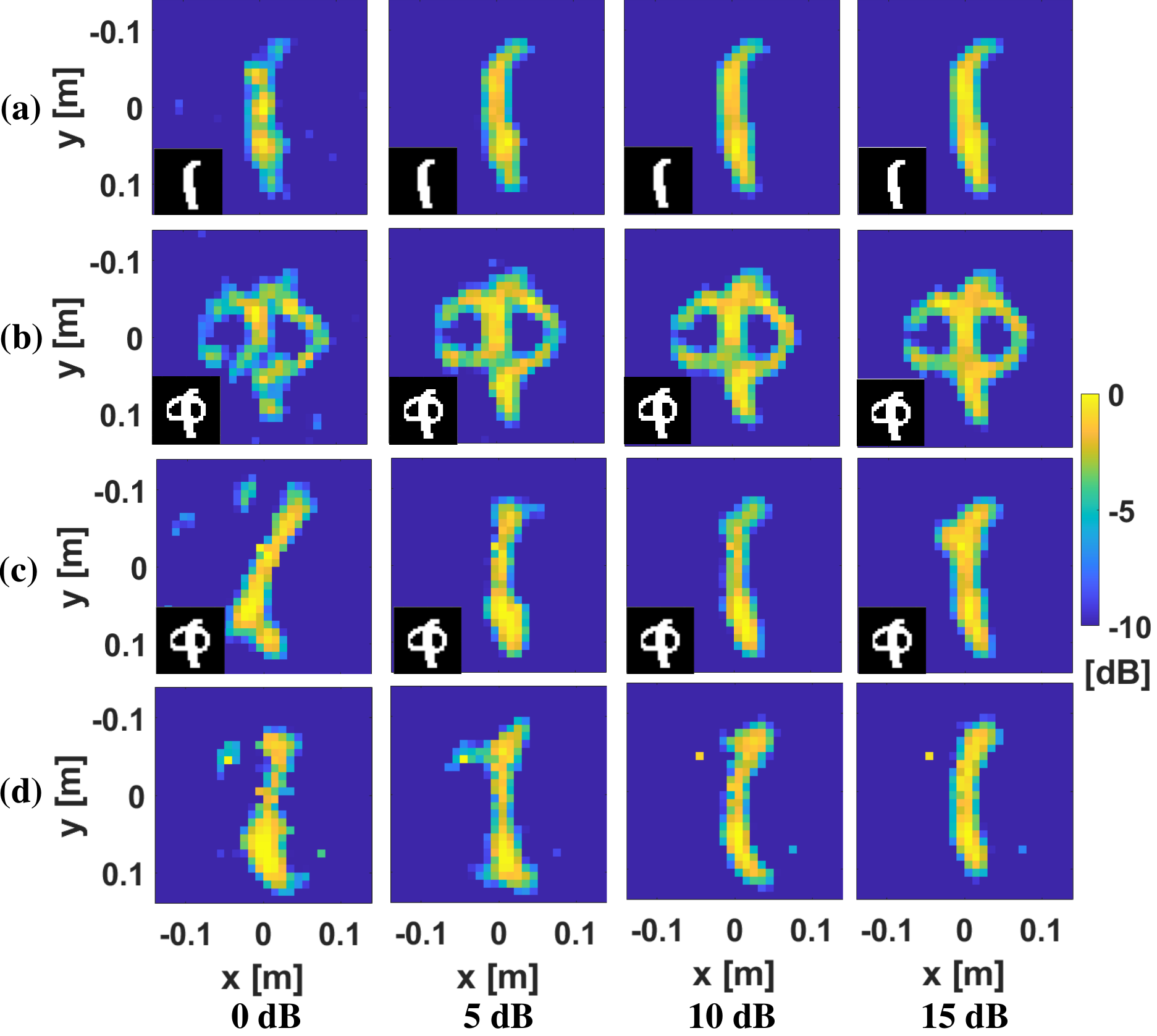}
    \caption{Comparisons of image reconstructions are presented using: (a) the conventional method described in (\ref{eq2_ls}) applied to back-scattered measurements from an unobstructed target of interest; (b) the same conventional method applied to measurements from an obstructed target; (c) the proposed cGAN-STM approach applied to measurements from the obstructed target; and (d) the cGAN-STM-Lite model applied to measurements from the obstructed target. {\color{black}The colour scale represents normalized reflectivity in dB. The maximum (1) is displayed at $0$ dB, and values below $-10$ dB are clipped.}}
    \label{fig:noise}
\end{figure}

Representative reconstructed images are shown in Figure \ref{fig:noise}. Specifically, Figs. \ref{fig:noise}(a) and (b) present the reconstructions of unobstructed and obstructed targets by using the conventional method outlined in (\ref{eq2_ls}), respectively, while Fig. \ref{fig:noise}(c) shows the cGAN-STM reconstructions for obstructed targets. Although performance metrics degrade significantly at very low SNRs, the target outlines remain visually recognizable.

\begin{table}[htbp]
\centering
\caption{Comparison of cGAN-STM Performance under Different SNR Levels.}
\setlength{\tabcolsep}{14.6pt}
\begin{tabular*}{\columnwidth}{|c|c|c|c|c|}
\hline
\textbf{SNR (dB)}  & \textbf{0}   & \textbf{5}   &   \textbf{10}   &  \textbf{15}   \\ \hline
\textbf{NMSE}      & 0.565  & 0.412 & 0.278 & 0.168\\ \hline
\textbf{SSIM}      & 0.183  & 0.322 & 0.513 & 0.700  \\ \hline
\end{tabular*}
\label{tab:noise}
\end{table}

\begin{figure}
    \centering
    \includegraphics[width=\columnwidth]{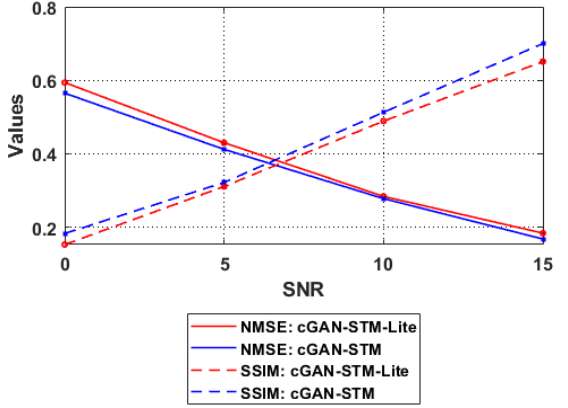}
    \caption{{\color{black}Average NMSE and SSIM of reconstructed images obtained by cGAN-STM and cGAN-STM-Lite under different SNR levels.}}
    \label{fig:noise_me}
\end{figure}

{\color{black}
In addition, a robustness analysis under noisy conditions is conducted to further investigate the cause of the performance degradation. In particular, Figs.~\ref{fig:noise}(c) and (d) compare reconstructions obtained by cGAN-STM and cGAN-STM-Lite, clearly demonstrating the superior performance of cGAN-STM under noisy conditions. As illustrated in Fig.~\ref{fig:noise_me}, with increasing SNR, NMSE decreases and SSIM increases for both models. However, across all noise levels, cGAN-STM consistently achieves lower NMSE and higher SSIM than cGAN-STM-Lite, indicating that the STM does not introduce additional sensitivity to noise.

Since STM is not the source of the degradation, the observed noise sensitivity can be attributed mainly to the inherent limitations of GANs, which are known to be unstable to input perturbations \cite{kos2018adversarial, creswell2018generative, yuan2019adversarial}. In light of this, incorporating noise-aware modules or physics-informed priors with the proposed cGAN-STM will be an important direction for future work to further enhance robustness against noise.
}

\begin{figure}[htb]
    \centering
    \includegraphics[width=\columnwidth]{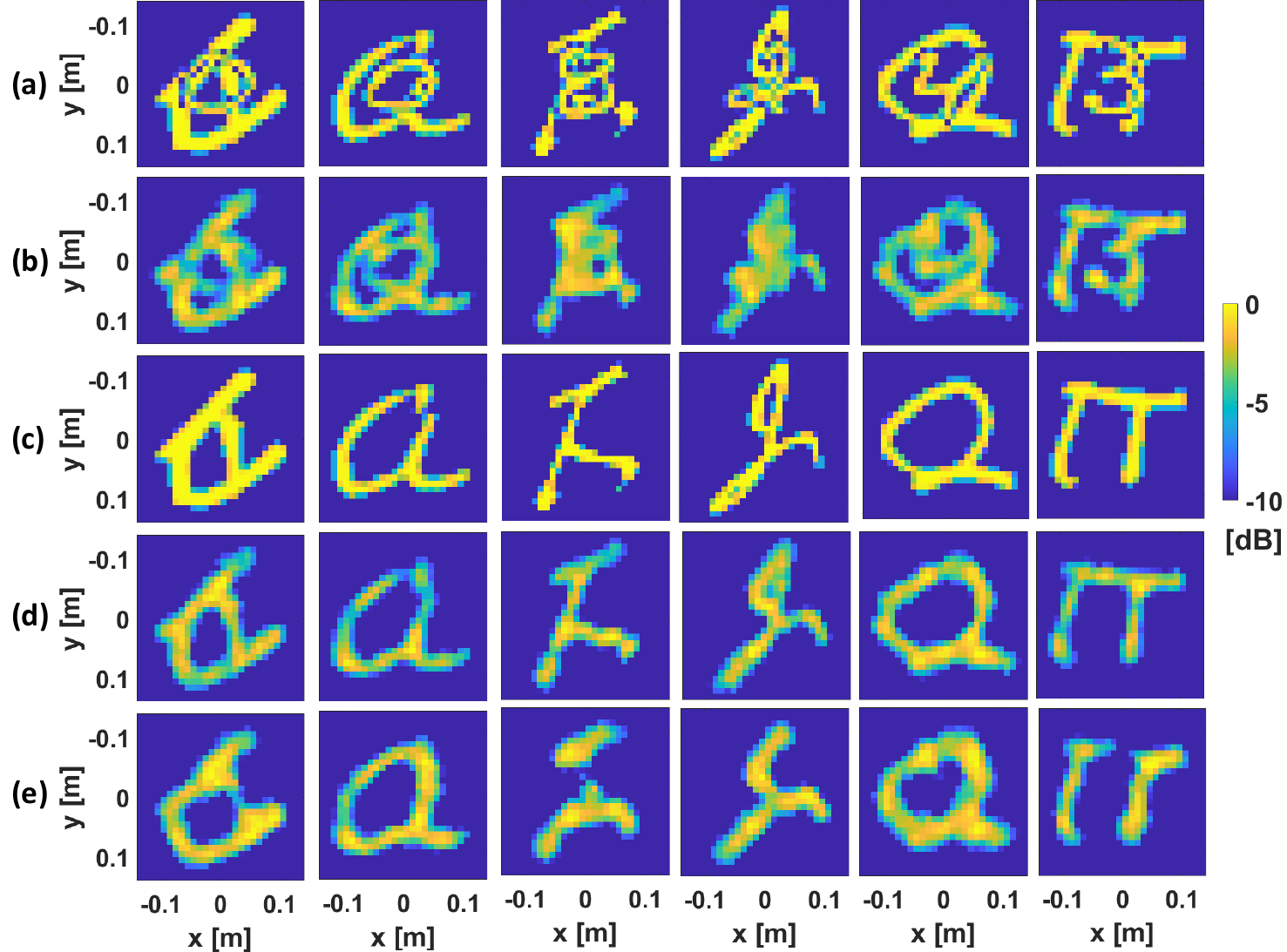}
    \caption{Comparison of image reconstructions of the obstructed imaging targets based on numerically synthesized back-scattered measurements using the experimentally measured sensing matrix: (a) obstructed imaging targets; (b) the image reconstructions of (a) retrieved using conventional method outlined in (\ref{eq2_ls}); (c) the unobstructed targets of interest; (d) the image reconstructions of (c) obtained using the conventional method; and (e) the image reconstructions using the back-scattered measurements of obstructed targets shown in (a), obtained by the proposed cGAN-STM. {\color{black}The colour scale represents normalized reflectivity in dB. The maximum (1) is displayed at $0$ dB, and values below $-10$ dB are clipped.}}
    \label{fig:reversal}
\end{figure}

\begin{figure}[htb]
    \centering
    \includegraphics[width=\linewidth]{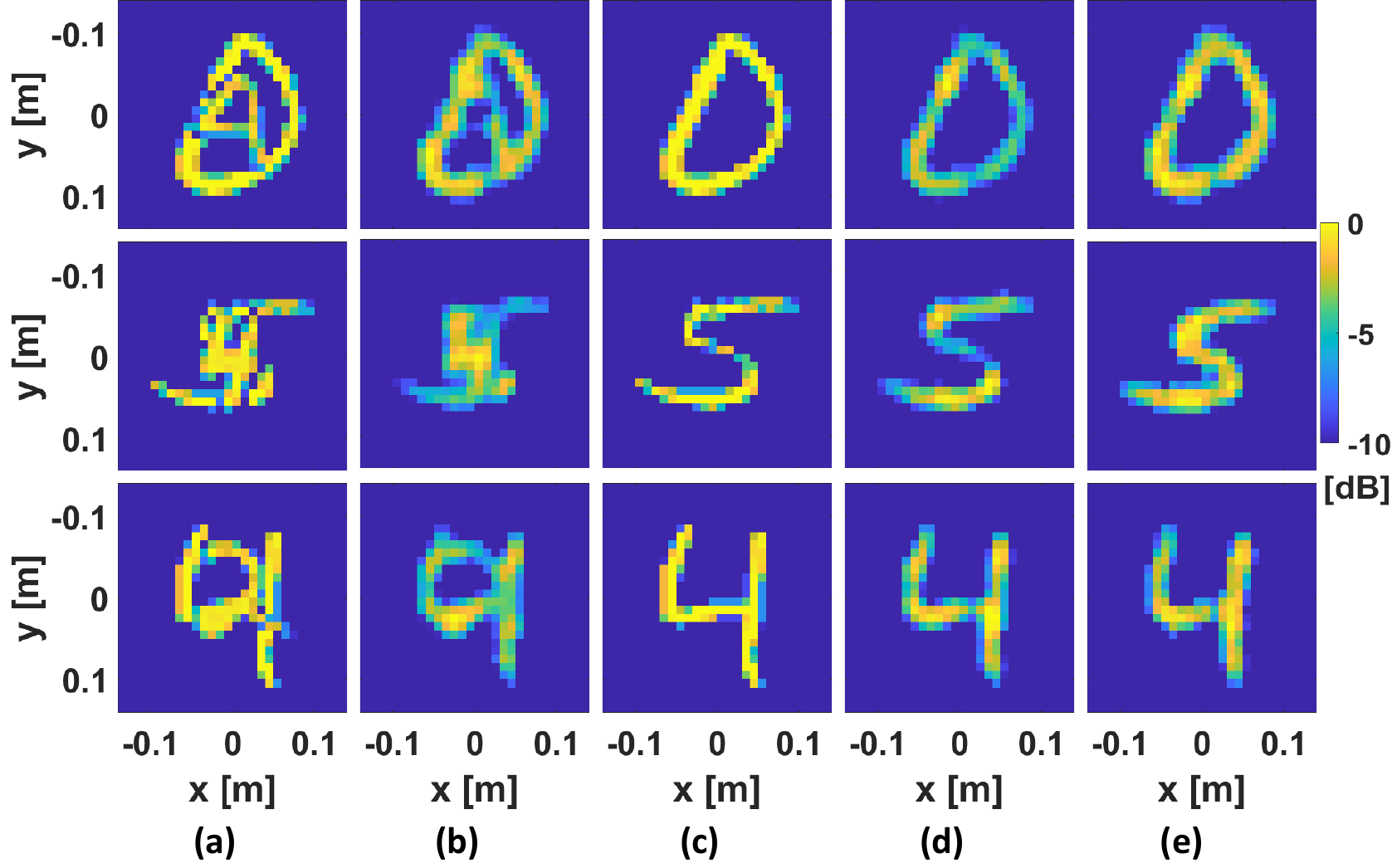}
    \caption{{\color{black}Comparison of image reconstructions of the obstructed imaging targets based on numerically synthesized back-scattered measurements using the experimentally measured sensing matrix: (a) obstructed imaging targets; (b) the image reconstructions of (a) retrieved using conventional method outlined in (\ref{eq2_ls}); (c) the unobstructed targets of interest; (d) the image reconstructions of (c) obtained using the conventional method; and (e) the image reconstructions using the back-scattered measurements of obstructed targets shown in (a), obtained by the proposed cGAN-STM. The colour scale represents normalized reflectivity in dB. The maximum (1) is displayed at $0$ dB, and values below $-10$ dB are clipped.}}
    \label{fig:osamet}
\end{figure}

\begin{figure}[htb]
    \centering
    \includegraphics[width=\linewidth]{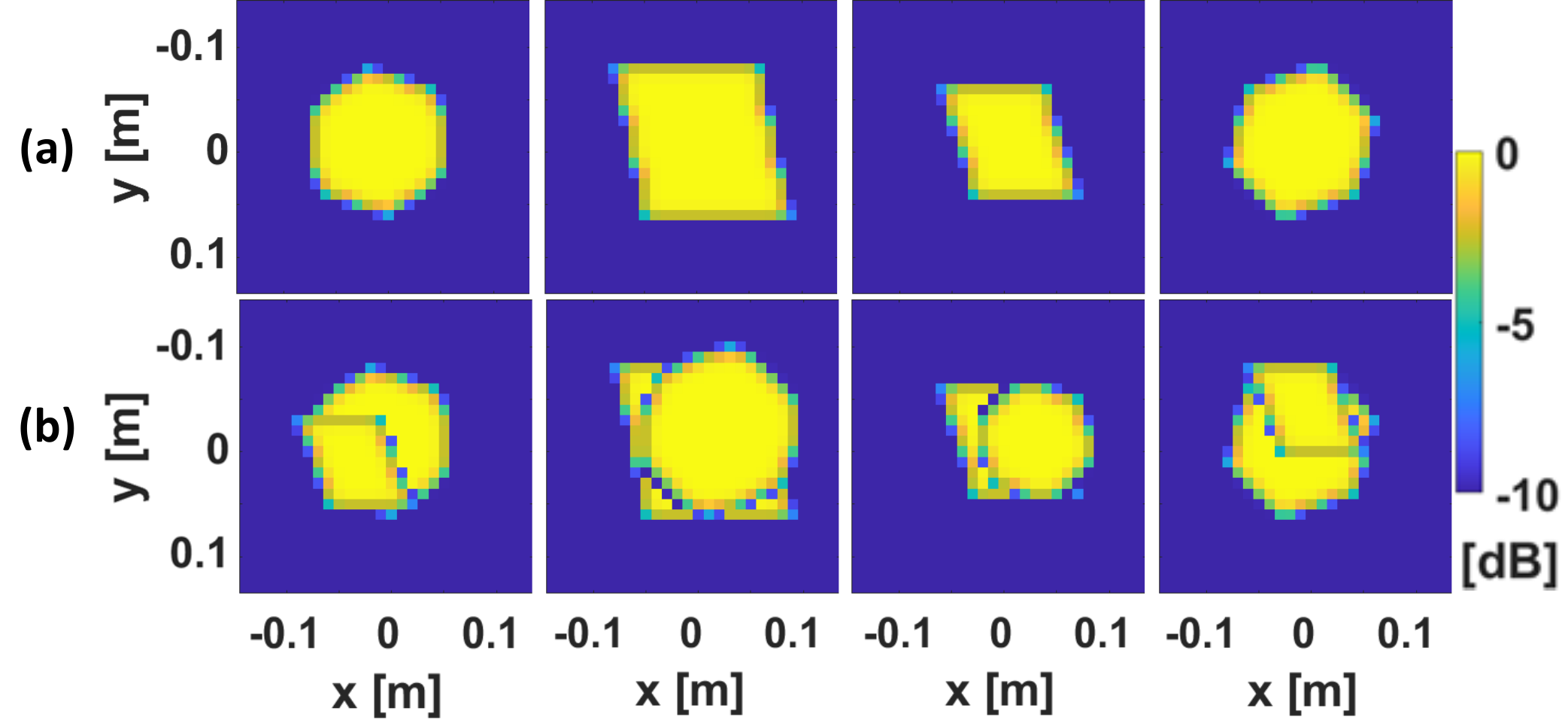}
    \caption{{\color{black}Examples of targets used for {\color{black}calibration}: (a) The  targets of interest (unobstructed) and (b) the obstructed targets. The colour scale represents normalized reflectivity in dB. The maximum (1) is displayed at 0 dB, and values below -10 dB are clipped.}}
    \label{fig:ft_target}
\end{figure}

\begin{figure}[htb]
    \centering
    \includegraphics[width=\linewidth]{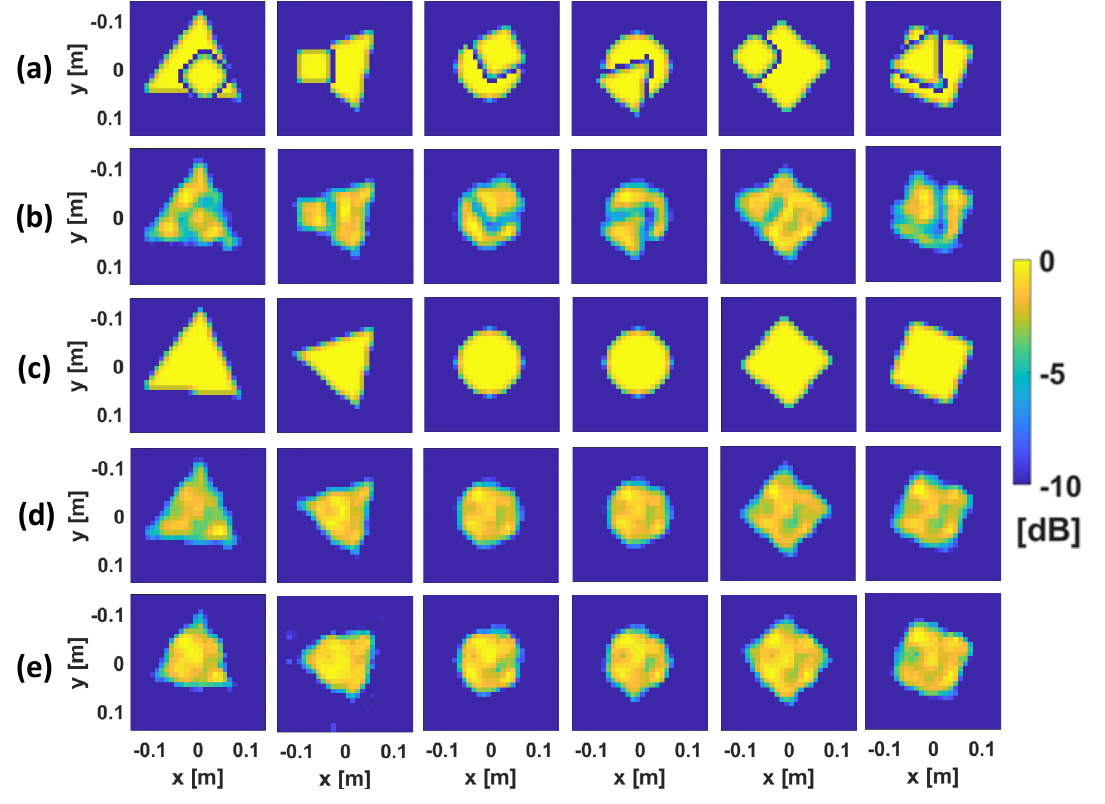}
    \caption{Comparison of image reconstructions of the obstructed imaging targets based on numerically synthesized back-scattered measurements using the experimentally measured sensing matrix: (a) obstructed imaging targets; (b) the image reconstructions of (a) retrieved using the conventional method outlined in (\ref{eq2_ls}); (c) the targets of interest of (a); (d) the image reconstructions of (c) obtained by using the conventional method; and (e) the image reconstructions using the back-scattered measurements of obstructed targets shown in (a), obtained by the {\color{black}calibrated} cGAN-STM. The colour scale represents normalized reflectivity in dB. The maximum (1) is displayed at 0 dB, and values below -10 dB are clipped.}
    \label{fig:ExperimentAfterFinetune}
\end{figure}

\begin{figure*}[htb]
    \centering
    \includegraphics[width=\linewidth]{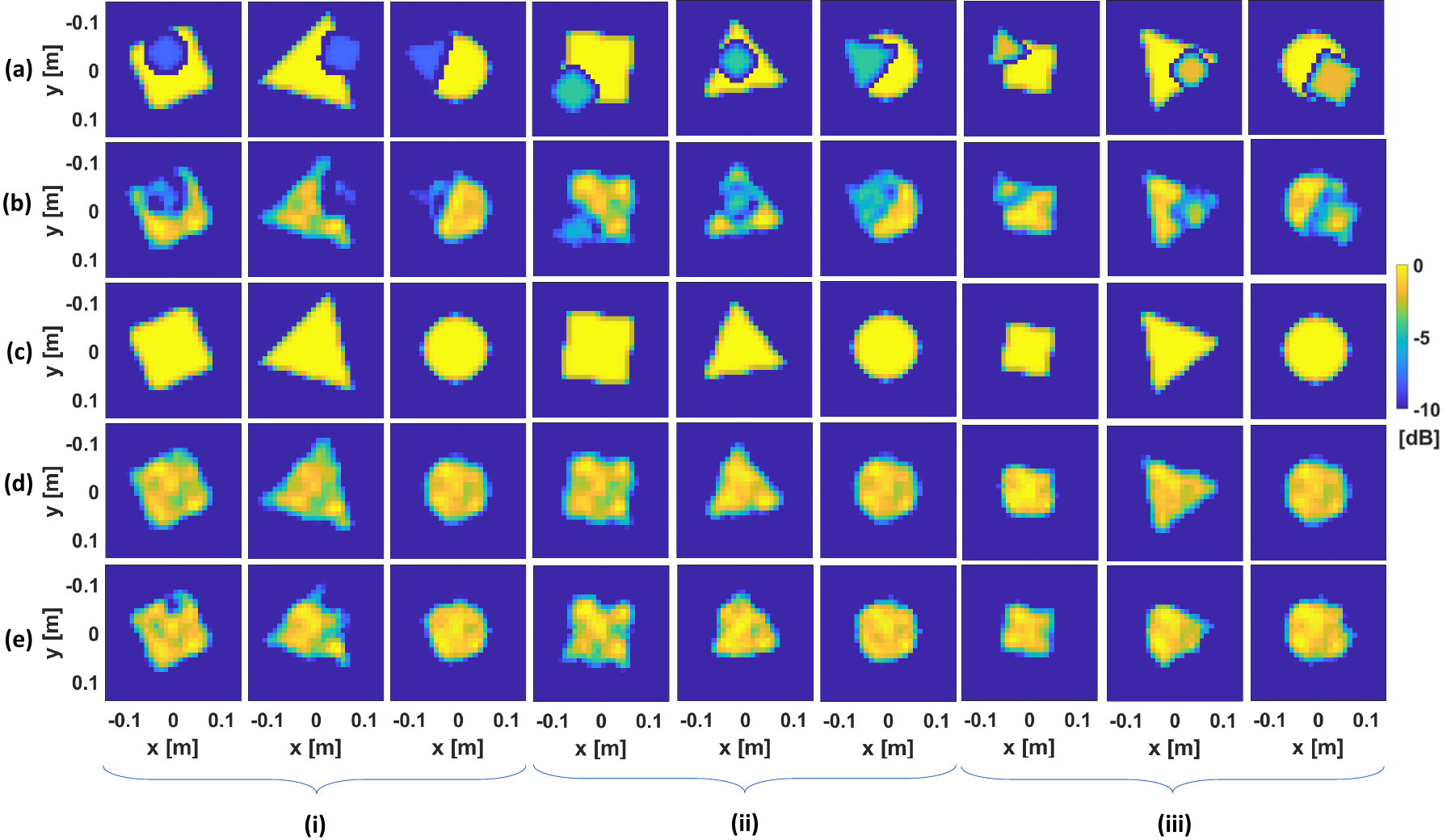}
    \caption{{\color{black}Comparison of image reconstructions of the obstructed imaging targets with reflectivity levels of (i) 0.4, (ii) 0.6, and (iii) 0.8. Each group (i)–(iii) corresponds to a different reflectivity level, and within each group: (a) obstructed imaging targets; (b) the image reconstructions of (a) retrieved using the conventional method outlined in (\ref{eq2_ls}); (c) the targets of interest of (a); (d) the image reconstructions of (c) obtained by using the conventional method; and (e) the image reconstructions using the back-scattered measurements of obstructed targets shown in (a), obtained with the proposed cGAN-STM after additional training. The colour scale represents normalized reflectivity in dB. The maximum (1) is displayed at 0 dB,  and values below -10 dB are clipped. }}
    \label{fig:Different_Reflectivity}
\end{figure*}

{\color{black}
\subsection{Analysis of Occlusion Suppression with Structurally Similar and Previously Learned Targets}
\label{sec:rev}
To analyze the model’s suppression behavior when previously learned target classes act as occluders, two related experiments are conducted.
{\color{black}
\subsubsection{Target-Occluder Role Reversal Test}
In this setting, E-MNIST letters are used as the targets of interest, while MNIST digits serve as obstructing objects. In other words, the databases for the targets and occluders are swapped (from MNIST to E-MNIST and from E-MNIST to MNIST, respectively) compared with the previous studies.
Since MNIST digits were used as targets during training, reusing them as occluders enables testing for potential memorization effects. This experiment therefore serves as a sanity check against overfitting and assesses the model’s robustness to familiar but irrelevant occlusions.
Following the data generation procedure in Section~\ref{sec: Data Generation}, 1000 unseen letters and digits are randomly selected from the MNIST and E-MNIST testing datasets, respectively. Each letter is obstructed by a digit, resulting in 1000 back-scattered measurements.

Six representative examples are shown in Fig.~\ref{fig:reversal}. Figs.~\ref{fig:reversal}(a) and \ref{fig:reversal}(c) present the obstructed and unobstructed targets, and Figs.~\ref{fig:reversal}(b) and \ref{fig:reversal}(d) show their corresponding reconstructions using the conventional method described in (\ref{eq2_ls}). Fig.~\ref{fig:reversal}(e) illustrates the reconstruction results obtained by the proposed cGAN-STM from the back-scattered measurements in Fig.~\ref{fig:reversal}(a). The average NMSE and SSIM are 0.171 and 0.615, respectively.

\subsubsection{Same-category Occlusion Test}
To further examine the model’s performance when both targets and occluders belong to the same domain, another experiment is conducted using only the MNIST dataset. A total of 1000 digits are randomly selected as targets and another 1000 as occluders, both from the MNIST testing set. Following the same data generation process, 1000 back-scattered measurements of obstructed digits are obtained.

Three representative examples are shown in Fig.~\ref{fig:osamet}. Figs.~\ref{fig:osamet}(a) and (c) display the obstructed and unobstructed targets, and Figs.~\ref{fig:osamet}(b) and (d) present their reconstructions using the conventional method described in (\ref{eq2_ls}). Fig.~\ref{fig:osamet}(e) shows the reconstructions obtained by the proposed cGAN-STM using the back-scattered measurements from Fig.~\ref{fig:osamet}(a). The average NMSE and SSIM are 0.068 and 0.871, respectively.

Overall, the results indicate that the proposed cGAN–STM also exhibits adaptive suppression behavior toward previously learned target classes, effectively reducing non-target featuress even when the occluders originate from familiar categories.

\subsection{Generalization and Adaptability Analysis}
\label{Sec:Adaptability Analysis}
To investigate the adaptability of the proposed cGAN-STM, experiments are conducted using unseen canonical-shaped targets. Specifically, three geometric shapes, namely disk, square, and triangle, are considered as the basis for constructing the obstructed targets. Each shape includes 150 randomly rotated and resized versions. For each obstructed target, the target of interest is selected from one of the three shapes, while the obstruction is chosen from the remaining two. The same experimentally measured sensing matrix and imaging configuration are used to ensure consistency with the previous setup. The geometric testing dataset is generated following the procedure described in  Section \ref{sec: Data Generation}. 

To improve reconstruction quality, the previously trained cGAN-STM is further {\color{black}calibrated} using a {\color{black}calibrated} dataset consisting of a limited amount of additional data. In the {\color{black}calibrated} dataset, both the imaging targets and the obstructions are randomly selected from parallelogram and hexagon shapes, with representative examples shown in Fig. \ref{fig:ft_target}. Fig. \ref{fig:ft_target}(a) shows the targets of interest, while Fig. \ref{fig:ft_target}(b) shows the related obstructed targets. The {\color{black}calibrated} dataset is also generated following the steps in Section \ref{sec: Data Generation}. The {\color{black}calibrated} dataset contains 300 samples in total. The cGAN-STM was {\color{black}calibrated} for 10 epochs until the loss converged, using a learning rate of  $1\times10^{-5}$, which is one-tenth of the original learning rate used in the initial training.

Representative reconstruction results are presented in Fig. \ref{fig:ExperimentAfterFinetune} to qualitatively illustrate the typical reconstruction behavior observed across the test samples. Figs. \ref{fig:ExperimentAfterFinetune}(a) and (c) show the obstructed and unobstructed targets, and Figs. \ref{fig:ExperimentAfterFinetune}(b) and (d) present the reconstructed images using the conventional method (\ref{eq2_ls}). Fig. \ref{fig:ExperimentAfterFinetune}(e) presents the image reconstructions using the {\color{black}calibrated} cGAN-STM and the back-scattered measurements from Fig. \ref{fig:ExperimentAfterFinetune}(a). Comparing Fig. \ref{fig:ExperimentAfterFinetune}(e) with Fig. \ref{fig:ExperimentAfterFinetune}(b) reveals that the proposed cGAN-STM significantly suppresses non-target features introduced by the obstructing targets, particularly along object edges. Quantitatively, the average NMSE and SSIM values of the geometric testing dataset are 0.071 and 0.854, respectively.  

To further evaluate the performance of the {\color{black}calibrated} cGAN-STM under diverse physical scattering conditions, three additional numerical datasets are generated. These datasets are based on similar canonical geometries but incorporate variations in reflectivity, target size, and orientation to emulate more realistic CMI scenarios. The representative reconstruction results are shown in Figs. \ref{fig:Different_Reflectivity}(i), (ii), and (iii). These images present the results when the obstructions have reflectivity levels of 0.4, 0.6 and 0.8, respectively. The obstructed targets are displayed in Fig. \ref{fig:Different_Reflectivity}(a) whilst the corresponding targets of interest are depicted in Fig. \ref{fig:Different_Reflectivity}(c). The image reconstructions of Figs. \ref{fig:Different_Reflectivity}(a) and \ref{fig:Different_Reflectivity}(c) obtained with the conventional method are shown in Figs. \ref{fig:Different_Reflectivity}(b) and (d), respectively. Fig. \ref{fig:Different_Reflectivity}(e) presents the image reconstructions using the {\color{black}calibrated} cGAN-STM and the back-scattered measurements from Fig. \ref{fig:Different_Reflectivity}(a). Additionally, the average metric values are shown in Table \ref{tab:diff_ref}.

\begin{table}[htb]
\centering
\caption{{\color{black}Average metric values between reconstructed images using the proposed cGAN-STM and conventional methods when the target of interest has different reflectivity.}}
\setlength{\tabcolsep}{19.7pt}
\begin{tabular*}{\columnwidth}{|c|c|c|c|}
\hline
\textbf{Reflectivity}  & \textbf{0.4} & \textbf{0.6} & \textbf{0.8} \\ \hline
\textbf{NMSE} & 0.143        & 0.112        & 0.088        \\ \hline
\textbf{SSIM} & 0.798        & 0.811        & 0.833        \\ \hline
\end{tabular*}
\label{tab:diff_ref}
\end{table}

In addition, a full training baseline is implemented, where the model is trained from scratch using the same 300-sample calibration dataset as used in the lightweight adaptation. The full training baseline achieves an NMSE of 0.099 and an SSIM of 0.773 on the geometric testing dataset. Under this limited data calibration setting, the proposed lightweight adaptation achieves improved reconstruction performance compared with full training from scratch, indicating that the pretrained network can be effectively adapted to previously unseen geometric shapes using only a small calibration dataset.

These results suggest that the proposed cGAN-STM can be efficiently adapted to new domains through lightweight calibration with a small number of samples. This adaptation process can be interpreted as a calibration step, where the model is adjusted to a new measurement environment using a limited set of representative samples. Furthermore, the results indicate that the network learns transferable representations rather than memorizing specific target or obstruction shapes, enabling efficient adaptation to previously unseen configurations.

\subsection{Limitation and Future Work}
\label{Sec:Limitation and Future Work}
Although the proposed cGAN–STM demonstrates strong effectiveness in reconstructing obstructed targets, its performance may degrade under extreme conditions. These include highly noisy environments, excessively large occlusions, or scenarios where obstructions dominate the feature space, leading to reduced suppression selectivity. Future work will aim to enhance the STM’s robustness and adaptability under such conditions through improved feature modulation, attention-guided suppression, or physics-informed learning strategies.

Also, the datasets used in this study also impose constraints on physical realism. While the MNIST and E-MNIST datasets are well suited for demonstrating the feasibility of the proposed method, they provide a simplified representation of the CMI problem. Realistic scenarios involve targets and clutter with continuous-valued reflectivity distributions, complex 3D geometries. Future work will consider complex targets with varying material properties to further optimize and validate the proposed model.

\section{Conclusion}
\label{sec: Conclusion}
This article proposed a novel cGAN integrated with deep learning based soft-threshold modules, called cGAN-STM, to address the image reconstruction problem in CMI scenarios involving obstructed imaging targets. The proposed cGAN-STM learns the information related to the targets of interest from the back-scattered measurements of the obstructed targets, generating the image reconstructions of the targets of interest. The model was evaluated with 10000 testing sample sets, where the obtained key image reconstruction metrics, namely NMSE and SSIM, were 0.066 and 0.876, respectively. Experimental tests further demonstrated the feasibility of the approach in practical environments. Comparisons with other state-of-the-art learning models in the literature confirmed the advantages of cGAN-STM. Additional experiments, including {\color{black}STM mechanism analysis}, variations in occluder size, target–occluder role reversal, different SNR noise conditions, {\color{black}and adaptability on unseen geometric shaped targets, showed its effectiveness and potential scalability to more complex targets.}

By means of this approach, the need for the image reconstruction step and the required manual work for removing the obstruction can be essentially eliminated. This approach offers the potential to enhance the overall efficiency of CMI, which can be particularly useful in some contexts and applications, such as security screening.

\bibliographystyle{IEEEtran}
\bibliography{bib.bib}
\end{document}